\documentclass[10pt]{article}

\usepackage{a4wide}
\usepackage{amssymb}
\usepackage{mathrsfs}

\title{The locally covariant Dirac field}
\author{Ko Sanders\thanks{E-mail:
jacobus.sanders@theorie.physik.uni-goe.de}\\
Institute of Theoretical Physics\\
University of G\"ottingen,\\
Friedrich-Hund-Platz 1, D-37077 G\"ottingen\\
and\\
Courant Research Centre\\
''Higher Order Structures in Mathematics'',\\
University of G\"ottingen}
\date{November 9, 2009}

\newtheorem{definition}{Definition}[section]
\newtheorem{theorem}[definition]{Theorem}
\newtheorem{proposition}[definition]{Proposition}
\newtheorem{corollary}[definition]{Corollary}
\newtheorem{lemma}[definition]{Lemma}
\newtheorem{remark}[definition]{Remark}
\newenvironment{proof*}{\smallskip\par\noindent\emph{Proof. }
 \ignorespaces}{\hfill$\Box$\smallskip\par\ignorespaces}

\newcommand{\dirop}{\ensuremath{\nabla\!\!\!\!\!\!\;/ \,}}
\newcommand{\map}[3]{\ensuremath{#1\!:\!#2\!\rightarrow\!#3}}
\newcommand{\id}{\ensuremath{\mathrm{id}}}
\newcommand{\fsl}[1]{\ensuremath{#1\!\!\!\!\!\;/ \,}}
\newcommand{\cat}[1]{\ensuremath{\mathfrak{#1}}}
\newcommand{\func}[1]{\ensuremath{\mathbf{#1}}}
\newcommand{\nt}[3]{\ensuremath{#1\!:\!#2\!\Rightarrow\!#3}}
\newcommand{\nte}[3]{\ensuremath{#1\!:\!#2\!\Leftrightarrow\!#3}}
\newcommand{\alg}[1]{\ensuremath{\mathcal{#1}}}
\newcommand{\Test}{\ensuremath{C^{\infty}}}

\newcommand{\Z}{\ensuremath{\mathbb{Z}}}
\newcommand{\R}{\ensuremath{\mathbb{R}}}
\newcommand{\C}{\ensuremath{\mathbb{C}}}

\begin{document}

\maketitle

\begin{abstract}
We describe the free Dirac field in a four dimensional spacetime as a locally
covariant quantum field theory in the sense of Brunetti, Fredenhagen and Verch,
using a representation independent construction. The freedom in the geometric
constructions involved can be encoded in terms of the cohomology of the
category of spin spacetimes. If we restrict ourselves to the observable algebra
the cohomological obstructions vanish and the theory is unique. We establish
some basic properties of the theory and discuss the class of Hadamard states,
filling some technical gaps in the literature. Finally we show that the
relative Cauchy evolution yields commutators with the stress-energy-momentum
tensor, as in the scalar field case.
\end{abstract}

\section{Introduction}\label{sec_intro}

Quantum field theory in curved spacetime is relevant for several purposes,
such as the construction of cosmological models and to obtain a better
understanding of quantum field theory in Minkowski spacetime. In order to
achieve this goals in a more realistic setting it is important to go beyond
the well-studied free scalar field. In this paper we will present a proof,
already contained in \cite{KS2}, of the fact that the free Dirac field in a
four dimensional globally hyperbolic spacetime can be described as a locally
covariant quantum field theory in the sense of \cite{Brunetti+}.

Our presentation of the Dirac field is representation independent and we
emphasise categorical methods throughout in order to point out an interesting
problem concerning the uniqeness of the theory. The obstruction for the
definition of a unique theory can be formulated in terms of the cohomology of
the category of spacetimes with a spin structure, in particular its first
Stiefel-Whitney class. It seems difficult to compute this class for a category,
but  we will show that a unique theory can always be obtained by restriction
to the observable algebrass generated by even polynomials in the field, in
which case the cohomological obstructions vanish.

Hadamard states can be defined in terms of a series expansion of their
two-point distribution, detailing their local singularity structure.
Alternatively, they can be characterised by a microlocal condition. The
equivalence of these two definitions has been investigated by several
authors using different techniques of proof, but in our opinion none of
these arguments has been fully convincing. In our discussion we hope to
close any remaining gaps in the different proofs and establish the
equivalence on firm ground.

We also compute the relative Cauchy evolution of this field and obtain
commutators with the stress-energy-momentum tensor, in complete analogy
with the scalar field case (\cite{Brunetti+}). For this we use a
point-splitting procedure to renormalise the stress-energy-momentum tensor.
Because we only need commutators with this tensor we do not need to treat
the so-called trace anomaly, a divergent multiple of the identity operator,
in detail. We refer the interested reader to \cite{Dappiaggi+}, who also
construct the extended algebra of Wick powers, relevant for perturbation
theory.

The contents of this paper are organised as follows. In section
\ref{sec_math} we review some of the mathematical background material
that we need in order to describe the Dirac field. This includes first of
all the Dirac algebra and the Spin group, followed by a categorical
formulation of some of the differential geometry that we will need. In
section \ref{sec_CLDirac} we describe the classical free Dirac field,
starting with the geometric and algebraic aspects in subsections
\ref{ssec_Dgeom} and \ref{ssec_Dstruc} and the equations of motion and
their fundamental solutions in subsection \ref{ssec_Deqn}. We discuss
the uniqueness of the functorial constructions and their cohomological
obstructions in subsection \ref{ssec_uniqueness}. We then proceed to
the quantum Dirac field in section \ref{sec_LCDirac}. In subsection
\ref{ssec_quant} we quantise the classical Dirac field in a local and
covariant way and collect some of its basic properties. Subsection
\ref{ssec_Had} deals with Hadamard states and includes a discussion of
the existing results concerning the equivalence of the microlocal and the
series expansion definitions. For this purpose we also refer to appendix
\ref{sec_ma}, which contains several relevant and useful (but expected)
results in microlocal analysis. Subsection \ref{ssec_RCE} contains our
discussion of the relative Cauchy evolution of the free Dirac field,
obtaining commutators with the stress-energy-momentum tensor, but the
proof of our main result there is deferred to appendix \ref{sec_proof},
because it consists of rather involved computations. Finally we end with
some conclusions.

Our presentation of locally covariant quantum field theory is based on the
original \cite{Brunetti+} and on \cite{Fewster}. For the Dirac field in
curved spacetime we largely follow \cite{Dimock} and \cite{Fewster+}, as
well as our earlier \cite{KS2}. For results on Clifford algebras we
refer to \cite{Lawson+} (see also \cite{Coquereaux} for a short review).

\section{Mathematical preliminaries}\label{sec_math}

To prepare for our discussion of the locally covariant Dirac field
we present in the current section some mathematical preliminaries
concerning the Dirac algebra, the Spin group and a categorical
formulation of relevant aspects of differential geometry. These
merely serve to fix our notation and set the scene for the subsequent
sections. We also point out the relations with some other
definitions and conventions in the literature.

\subsection{The Dirac algebra and the Spin group}\label{ssec_spin}

The Spin group can be embedded in the Clifford algebra of Minkowski
spacetime, which we call the Dirac algebra. Therefore we will first
briefly recall some results on Clifford algebras, for wich we refer
to \cite{Lawson+} (note the difference in sign convention in the
Clifford multiplication).

Let $\mathbb{R}^{r,s}$ be a finite dimensional real vector space
with dimension $n=r+s$ and with a non-degenerate bilinear form
$g_{ab}$ which has $r$ positive and $s$ negative eigenvalues. The
\emph{Clifford algebra} $Cl_{r,s}$ is defined as the $\R$-linear
associative algebra generated by a unit element $I$ and an
orthonormal basis $e_a$ of $\mathbb{R}^{r,n-r}$ subject to the
relations:
\[
e_ae_b+e_be_a=2g_{ab}I.
\]
This definition is independent of the choice of basis. We may identify
$\mathbb{R}^{r,s}\subset Cl_{r,s}$ as the subspace of monomials in the
basis $e_a$ of degree 1. The even, respectively odd, subspace of this
Clifford algebra is the one spanned by monomials of even, respectively
odd, degree in the basis vectors and is denoted by $Cl^0_{r,s}$,
respectively $Cl^1_{r,s}$. Note that the even subspace is also a
subalgebra. In the following we will be especially interested in
Minkowski spacetime, $M_0:=\mathbb{R}^{1,3}$, where the bilinear form
is $\eta=\mathrm{diag}(1,-1,-1,-1)$ and where we choose an orthonormal
basis $g_a$, $a=0,1,2,3$ with $\|g_0\|^2=1$, $\|.\|^2$ denoting the
Minkowski pseudo-norm squared. The associated Clifford algebra is
called the \emph{Dirac algebra} $D:=Cl_{1,3}$ and it is characterised
by
\begin{equation}\label{Clifford}
g_ag_b+g_bg_a=2\eta_{ab}I.
\end{equation}

As a vector space the Clifford algebra is naturally isomorphic to the
exterior algebra. This motivates the term \emph{volume form} for the
element $g_5:=g_0g_1g_2g_3$ (or in general $e:=e_1\cdots e_{r+s}$). Note
the following properties:
\begin{lemma}\label{CliffordProperties}
We have $g_5^2=-I$ and $g_5vg_5^{-1}=-v$ for all $v\in M_0$. More
generally, if $u\in M_0$ has $u^2=\|u\|^2I\not=0$, then
$u^{-1}=\frac{1}{\|u\|^2}u$ and $v\mapsto -uvu^{-1}$ defines a reflection
of $M_0$ in the hyperplane perpendicular to $u$.
\end{lemma}
\begin{proof*}
These equalities follow directly from (\ref{Clifford}). For the
last claim, e.g., we compute:
\[
-uvu^{-1}=v-(uv+vu)u^{-1}=v-\frac{2\langle u,v\rangle}{\|u\|^2}u,
\quad v\in M_0.
\]
\end{proof*}

Standard arguments with Clifford algebras \cite{Lawson+} give:
\[
D=Cl_{1,3}\simeq Cl^0_{1,4}\simeq Cl^0_{4,1},\quad
Cl_{4,1}\simeq M(4,\mathbb{C}),
\]
where $M(4,\mathbb{C})$ denotes the algebra of complex
$4\times 4$-matrices. In fact, $Cl_{4,1}$ is generated by the
generators $g_a$ of $D$ together with a central element $\omega$,
corresponding to $iI\in M(4,\mathbb{C})$. Hence:
\begin{equation}\label{subalgebra}
M(4,\mathbb{C})\simeq\mathbb{C}\otimes_{\mathbb{R}}D.
\end{equation}
This also implies that the center of $D$ is spanned by $I$ (over
$\mathbb{R}$). The following Fundamental Theorem provides all the
essential information we need on the Dirac algebra (for an
elementary algebraic proof we refer to Pauli \cite{Pauli}.):
\begin{theorem}[Fundamental Theorem]\label{fundamentaltheorem}
The Dirac algebra $D$ is simple and has a unique irreducible complex
representation (i.e.\ an $\R$-linear representation
$\map{\pi}{D}{M(n,\mathbb{C})}$), up to equivalence. This is the
representation $\map{\pi_0}{D}{M(4,\mathbb{C})}$ determined by
$\pi_0(g_a)=\gamma_a$ with the Dirac matrices
\[
\gamma_0:=\left(\begin{array}{cc}
O&I\\
I&0
\end{array}\right),\quad
\gamma_i:=\left(\begin{array}{cc}
O&-\sigma_i\\
\sigma_i&0
\end{array}\right),
\]
where $\sigma_i$ are the Pauli matrices
$\sigma_1:=\left(\begin{array}{cc}O&1\\ 1&0\end{array}\right)$,
$\sigma_2:=\left(\begin{array}{cc}O&-i\\ i&0\end{array}\right)$ and
$\sigma_3:=\left(\begin{array}{cc}1&0\\ 0&-1\end{array}\right)$.
The equivalence with another irreducible complex representation $\pi$ of
$D$ is implemented by $\pi(S)=L\pi_0(S)L^{-1}$ for all
$S\in D$, where $L\in GL(4,\mathbb{C})$ is unique up to a
non-zero complex factor.

Consequently, for every set of matrices $\gamma'_a\in M(4,\mathbb{C})$
satisfying equation (\ref{Clifford}) there is an
$L\in GL(4,\mathbb{C})$, unique up to a non-zero complex constant,
such that $\gamma'_a=L\gamma_aL^{-1}$.
\end{theorem}
\begin{proof*}
One can show \cite{Lawson+} that $D\simeq M(2,\mathbb{H})$, which is
simple because it is a full matrix algebra. The given matrices $\gamma_a$
satisfy the Clifford relations (\ref{Clifford}) and therefore extend to a
representation of $D$ in $M(4,\mathbb{C})$.

Any complex representation $\map{\pi}{D}{M(n,\mathbb{C})}$ extends to
a complex representation $\tilde{\pi}$ of $M(4,\mathbb{C})$, using
equality (\ref{subalgebra}) and and the trivial center of $D$, which is
irreducible if $\pi$ is irreducible. As
$M(4,\mathbb{C})$ has only one irreducible representation up to
equivalence (see \cite{Waerden}) this determines $\pi$ up to
equivalence, as stated. If $K,L\in GL(4,\mathbb{C})$ are two matrices
which implement the same equivalence, then $KL^{-1}$ commutes with $D$
and hence $K=cL$, where $c\in\mathbb{C}$ is non-zero because $K$ is
invertible. Note that $\pi'(g_a):=\gamma'_a$ extends to a complex
representation of $D$ in $M(4,\mathbb{C})$ which is faithful (as $D$ is
simple). The last statement then follows from the previous one.
\end{proof*}
For notational convenience we define $\gamma_5:=\pi_0(g_5)$.

We can define a determinant and trace function on $D$ by
$\det S=\det\pi(S)$ and $Tr(S)=Tr(\pi(S))$ for all $S\in D$, where
$\pi$ is any irreducible complex representation of $D$. This is
well-defined by the Fundamental Theorem. The following lemma is often
useful in computations:
\begin{lemma}\label{traceresults}
We have $Tr(g_ag_b)=4\eta_{ab}$ and
$Tr(\left[g_b,g_c\right]g_dg_a)=8(\eta_{cd}\eta_{ba}-\eta_{bc}\eta_{da})$.
\end{lemma}
\begin{proof*}
Using the cyclicity of the trace and the relations (\ref{Clifford}) we
find:
$Tr(g_ag_b)=\frac{1}{2}Tr(g_ag_b+g_bg_a)=Tr(\eta_{ab}I)=4\eta_{ab}$ and
\begin{eqnarray}
Tr([g_b,g_c]g_dg_a)&=&Tr(g_b[g_c,g_dg_a])=Tr(g_b\left\{g_c,g_d\right\}g_a-
g_bg_d\left\{g_c,g_a\right\})\nonumber\\
&=&2Tr(\eta_{cd}g_bg_a-g_bg_d\eta_{ca})=
8(\eta_{cd}\eta_{ba}-\eta_{bd}\eta_{ca}).\nonumber
\end{eqnarray}
\end{proof*}

We now turn to the Spin group, which is the universal double covering
group of the special Lorentz group and which can be constructed in an
elegant way inside the Dirac algebra.
\begin{definition}\label{def_spin}
The Pin and Spin groups of $Cl_{r,s}$ are defined as
\[
Pin_{r,s}:=\left\{S\in Cl_{r,s}|\ S=u_1\cdots u_k,\quad
u_i\in\mathbb{R}^{r,s},\quad u_i^2=\pm I\right\},
\]
\[
Spin_{r,s}:=Pin_{r,s}\cap Cl^0_{r,s}.
\]
We let $Spin^0_{1,3}$ denote the connected component of $Spin_{1,3}$
which contains the identity.

We also define the Lorentz group $\mathcal{L}:=O_{1,3}$, the special
Lorentz group $\mathcal{L}_+:=SO_{1,3}$ and the special ortochronous
Lorentz group $\mathcal{L}_+^{\uparrow}:=SO_{1,3}^0$, which is the
connected component of $\mathcal{L}_+$ containing the identity.
\end{definition}
The special ortochronous Lorentz group preserves the orientation and
time-orientation. For $S\in Pin_{1,3}$ the map $v\mapsto SvS^{-1}$ on
$M_0$ is a product of reflections (up to a sign) by Lemma
\ref{CliffordProperties}. Together with the fact that
$\det u=\|u\|^4$ for all $u\in M_0$ this gives rise to another useful
characterisation of the group $Pin_{1,3}$, which we shall not
prove:\footnote{The definition of the Spin group in \cite{Choquet+}
corresponds to our group $Pin_{1,3}$. In \cite{Dimock} and
$\cite{Fewster+}$ one uses the term Spin group for the group
\[
\mathcal{S}:=\left\{S\in M(4,\mathbb{C})| \det S=1,\quad SvS^{-1}\in M_0
\mathrm{\ for\ all\ } v\in M_0 \right\}.
\]
Note that this group cannot give a double covering of the Lorentz group,
as claimed in \cite{Dimock} (but not in \cite{Fewster+}), because for
any $S\in\mathcal{S}$ the matrices $iS,-S,-iS$ are in $\mathcal{S}$ too.
Its usefulness is based on its simple definition and the fact that
$\mathcal{S}^0=Spin^0_{1,3}$.}
\begin{proposition}\label{spineq}
$Pin_{1,3}=\left\{S\in D|\ \det S=1, \forall v\in M_0
SvS^{-1}\in M_0\right\}$.
\end{proposition}

It can be seen from Proposition \ref{spineq} that $Pin_{1,3}$ and
$Spin_{1,3}$ are indeed Lie groups. For the universal double covering
homomorphism $\Lambda$ between $Pin_{1,3}$ and the Lorentz group we
have the following formulae:\footnote{These results are well-known,
but we record them for definiteness to correct a sign error in the
spin connection (\ref{spinconnection}) that has occured in
\cite{Dimock,Fewster+,Dawson+}.}\footnote{Lower case Latin indices are
raised and lowered with $\eta^{ab}$, resp.\ $\eta_{ab}$ throughout.}
\begin{proposition}\label{covering}
The map $\map{\Lambda}{Pin_{1,3}}{\mathcal{L}}$ defined by
$S\mapsto \Lambda^a_{\ b}(S)\in M(4,\mathbb{R})$ such that
$Sg_bS^{-1}=g_a\Lambda^a_{\ b}(S)$ is the universal double covering
homomorphism of Lie groups, which restricts to the universal double
covering homomorphism
$Spin^0_{1,3}\rightarrow\mathcal{L}^{\uparrow}_+$. We have
$\Lambda^a_{\ b}(S)=\frac{1}{4}Tr(g^aSg_bS^{-1})$ and the inverse of the
derivative $\map{d\Lambda}{spin^0_{1,3}}{l_+^{\uparrow}}$ at $S=I$ is
given by:
\[
(d\Lambda)^{-1}(\lambda^b_{\ a})=\frac{1}{4}\lambda^b_{\ a}g_bg^a.
\]
\end{proposition}
\begin{proof*}
For the first sentence we refer to \cite{Lawson+} Theorem 2.10 and
subsequent remarks. Using the Clifford relations (\ref{Clifford})
we see that
\begin{eqnarray}
\Lambda^a_{\ b}(S)&=&\frac{1}{4}\eta^{ac}Tr(\eta_{cd}\Lambda^d_{\ b}(S)I)=
\frac{1}{8}\eta^{ac}Tr((g_cg_d+g_dg_c)\Lambda^d_{\ b}(S))\nonumber\\
&=&\frac{1}{4}\eta^{ac}Tr(g_cg_d\Lambda^d_{\ b}(S))=
\frac{1}{4}Tr(g^aSg_bS^{-1}).\nonumber
\end{eqnarray}
Expanding $\Lambda(S+\epsilon s+O(\epsilon^2))$ up to second order in
$\epsilon$ we find
$d\Lambda(s)^a_{\ b}=\frac{1}{4}Tr(\left[g_b,g^a\right]s)$.
We check that $L(\lambda^b_{\ a}):=\frac{1}{4}\lambda^b_{\ a}g_bg^a$ is
an inverse of $d\Lambda$:
\begin{eqnarray}
d\Lambda(L(\lambda^d_{\ e}))^a_{\ b}&=&\frac{1}{16}\eta^{ac}\eta^{ef}
\lambda^d_{\ e}Tr(\left[g_b,g_c\right]g_dg_f)=\frac{1}{2}\eta^{ac}
\eta^{ef}\lambda^d_{\ e}(\eta_{cd}\eta_{bf}-\eta_{bd}\eta_{cf})
\nonumber\\
&=&\frac{1}{2}(\lambda^a_{\ b}-\eta^{ae}\eta_{bd}\lambda^d_{\ e})
=\lambda^a_{\ b},\nonumber
\end{eqnarray}
where we used Lemma \ref{traceresults} and the symmetry properties of
$\lambda^d_{\ e}\in l_+^{\uparrow}$ in the last line.
\end{proof*}

\subsection{Some category theory and differential geometry}\label{ssec_geom}

The language of locally covariant quantum field theory uses category
theory to express the physical ideas of locality and covariance. Any
object or construction that is extended from a single spacetime (usually
Minkowski spacetime) to the categorical framework gets the adjective
''locally covariant''. The essence of local covariance seems to have a
geometric origin and, because the Dirac field in curved spacetimes
involves a substantial amount of geometric constructions, it will be
convenient to present the relevant differential geometry in a categorical
setting here. We refrain from the urge to call this ''locally covariant
differential geometry'', which appears to be a pleonasm.

A \emph{category} $\cat{C}$ consists of a set of objects $c$ and a
set of morphisms or arrows\footnote{It is very often convenient to
depict the morphisms in a diagram as arrows between objetcs.}
$\map{\gamma}{c_1}{c_2}$ between objects of $\cat{C}$, such that the
composition of morphisms, when defined, is associative and each
object admits an identity morphism (we refer to \cite{MacLane} for
more details). A \emph{(covariant) functor}
$\map{\func{F}}{\cat{C}}{\cat{B}}$ is a map between categories,
which maps objects $c$ to objects $\func{F}(c)$ and morphisms
$\map{\gamma}{c_1}{c_2}$ to morphisms
$\map{\func{F}(\gamma)}{\func{F}(c_1)}{\func{F}(c_2)}$ such that an
identity morphism maps to an identity morphism and the composition
of morphisms is preserved. A \emph{contravariant functor} 
$\map{\func{F}}{\cat{C}}{\cat{B}}$ is defined similarly, but
reverses the direction of the morphisms:
$\map{\func{F}(\gamma)}{\func{F}(c_2)}{\func{F}(c_1)}$. A
\emph{natural transformation}  $\nt{t}{\func{F}}{\func{G}}$ between
covariant functors $\map{\func{F}}{\cat{C}}{\cat{B}}$ and
$\map{\func{G}}{\cat{C}}{\cat{B}}$ is a map which assigns to each
object $c$ a morphism $t(c)$ of $\cat{B}$, called the
\emph{component} of $t$ at $c$, such that for every morphism
$\map{\gamma}{c_1}{c_2}$ of $\cat{C}$ we have
$t(c_2)\circ\func{F}(\gamma)=\func{G}(\gamma)\circ t(c_1)$, which
can be depicted as a commutative diagram. When a natural
transformation $t$ admits another natural transformation $s$ such
that $t(c)\circ s(c)=\id_c=s(c)\circ t(c)$ for all objects $c$,
then $t$ is called a \emph{natural equivalence}. In this case we
write $\nte{t}{\func{F}}{\func{G}}$. A natural transformation
between contravariant functors or between a covariant and a
contravariant functor is defined similarly, except that some arrows
in the commutative diagram are reversed.

A subcategory $\cat{B}$ of $\cat{C}$ consists of a subset of the
objects of $\cat{C}$ and a subset of its morphisms in such a way
that $\cat{B}$ still satisfies the axioms of a category. In our
case all categories will be concrete, i.e.\ the objects will be
sets with a certain structure and the morphisms will be maps
between sets. The identity morphism will always be the identity
map and the composition of maps, when defined, is automatically
associative. In short, our categories will be subcategories of
the category $\cat{Set}$, whose objects are sets\footnote{See
\cite{MacLane} for some relevant remarks concerning the
foundations of set theory and the use of small sets.} and whose
morphisms are maps.

For our discussion of differential geometry we start with the
following
\begin{definition}
The category $\cat{Man}^n$ of smooth manifolds is the category
whose objects are $\Test$ manifolds $\mathcal{M}$ of (finite)
dimension $n$ and whose morphisms are $\Test$ embeddings
$\map{\mu}{\mathcal{M}_1}{\mathcal{M}_2}$.

The category $\cat{Bund}'$ of fiber bundles is the category
whose objects are smooth fiber bundles 
$\map{p}{\mathcal{B}}{\mathcal{M}}$ over objects $\mathcal{M}$ of
$\cat{Man}^n$ with bundle projection map $p$, and whose morphisms
are $\Test$
maps $\map{\beta}{\mathcal{B}_1}{\mathcal{B}_2}$ covering a
morphism $\map{\mu}{\mathcal{M}_1}{\mathcal{M}_2}$ of
$\cat{Man}^n$, i.e.\ such that $p_2\circ\beta=\mu\circ p_1$. We
denote by $\cat{Bund}$ the subcategory whose morphisms restrict
to isomorphisms of the fibers.

The categories $\cat{VBund}_{\mathbb{R}}'$, respectively 
$\cat{VBund}_{\mathbb{C}}'$, of real (complex) vector bundles is
the subcategory of $\cat{Bund}'$ whose objects $\mathcal{V}$ are
real (complex) vector bundles and whose morphisms
$\map{\nu}{\mathcal{V}_1}{\mathcal{V}_2}$ are real (complex)
linear maps of the fibers. Again we denote by
$\cat{VBund}_{\mathbb{R}}$ and $\cat{VBund}_{\mathbb{C}}$ the
subcategories whose morphisms restrict to isomorphisms of the
fibers.
\end{definition}
We could have taken all smooth maps between manifolds as morphisms
of $\cat{Man}^n$ or allowed all dimensions. However,
local diffeomorphisms allow us to transport more structure,
which enables us to describe more of the canonical differential
geometric constructions as functors. We list the most important
examples below. For fiber bundles, on the other hand, it will be
useful to allow maps which are not isomorphisms on the
fibers.\footnote{The unprimed categories, whose morphisms are
isomorphisms of the fibers, can be described as fibered
categories over $\cat{Man}^n$, cf.\ \cite{MacLane+}
p.44.}\footnote{The functors $\map{\func{B}}{\cat{Man}^n}{\cat{Bund}'}$
below are all of a special type, namely they associate to a manifold
$\mathcal{M}$ a fiber bundle whose base space is again $\mathcal{M}$.
Although we will only use functors of this type when describing the
Dirac field, the restriction is not technically necessary in our
definitions.}
\begin{enumerate}
\item The functor
$\map{\func{T}}{\cat{Man}^n}{\cat{VBund}_{\mathbb{R}}}$
assigns to every manifold $\mathcal{M}$ the tangent bundle
$T\mathcal{M}$ and to every morphism
$\map{\mu}{\mathcal{M}_1}{\mathcal{M}_2}$ the differential
$\map{d\mu}{T\mathcal{M}_1}{T\mathcal{M}_2}$.
\item The functor\footnote{It is tempting to think of a
\emph{contravariant} functor that maps manifolds to their
cotangent bundles and morphisms $\mu$ to the pull-back,
$\mu^*\omega:=\omega\circ d\mu$, which indeed reverses the
directions of arrows and changes the order of compositions.
However, the pull-back is only defined on the image of $\mu$,
so in general this does not define a morphism in
$\cat{VBund}_{\mathbb{R}}'$.}
$\map{\func{T}^*}{\cat{Man}^n}{\cat{VBund}_{\mathbb{R}}}$ assigns
to every manifold $\mathcal{M}$ the cotangent bundle
$T^*\mathcal{M}$ and to every morphism
$\map{\mu}{\mathcal{M}_1}{\mathcal{M}_2}$ the push-forward
$\map{\mu_*}{T\mathcal{M}_1}{T\mathcal{M}_2}$, which is
defined as $\mu_*\omega:=\omega\circ d\mu^{-1}$.
\item Finite direct sums and tensor products of $\func{T}$ and
$\func{T}^*$ can also be described as functors, by extending
$d\mu$ and $\mu_*$ in the obvious way.
\item The functor
$\map{\func{\Lambda}^k}{\cat{Man}^n}{\cat{VBund}_{\mathbb{R}}}$
assigns to every manifold $\mathcal{M}$ the vector bundle
$\Lambda^k\mathcal{M}$ of exterior $k$-forms and to every
morphism $\mu$ the push-forward $\mu_*$ of such forms. Similarly
the functor
$\map{\func{\Lambda}}{\cat{Man}^n}{\cat{VBund}_{\mathbb{R}}}$
assigns to a manifold the exterior algebra and it maps morphisms
to push-forwards.
\item The functor
$\map{|\func{\Lambda}^n|}{\cat{Man}^n}{\cat{VBund}_{\mathbb{R}}}$
assigns to every spacetime $\mathcal{M}$ the one dimensional
trivial vector bundle of densities $|\Lambda^n\mathcal{M}|$,
where $n$ is the dimension of $\mathcal{M}$. This is the vector
bundle whose fiber at $x\in\mathcal{M}$ consists of functions
$\map{d}{\Lambda^n_x\mathcal{M}}{\mathbb{R}}$ such that
$d(r\omega)=|r|\omega$ for all $r\in\mathbb{R}$ and
$\omega\in\Lambda^n_x\mathcal{M}$ (cf.\ \cite{Baer+} appendix
A.3). A morphism $\mu$ is mapped to the push-forward defined by
$\mu_*d:=d\circ \mu^*$, where $\mu^*\omega:=\omega\circ d\mu$ is
the pull-back.
\item In general, for a functor
$\map{\func{V}}{\cat{Man}^n}{\cat{VBund}_{\mathbb{R}}}$ with
$\mathcal{M}\mapsto V\mathcal{M}$ and $\mu\mapsto\beta$, the dual
functor $\map{\func{V}^*}{\cat{Man}^n}{\cat{VBund}_{\mathbb{R}}}$
assigns to every manifold $\mathcal{M}$ the dual vector bundle
$V^*\mathcal{M}$ of $V\mathcal{M}$ and to each morphism
$\map{\mu}{\mathcal{M}_1}{\mathcal{M}_2}$ the push-forward morphism
$\beta_*$ defined by $\beta_*\omega:=\omega\circ\beta^{-1}$.
\item When $\map{\func{V}_i}{\cat{Man}^n}{\cat{VBund}_{\mathbb{R}}'}$
for $i=1,\ldots,n$ map $\mathcal{M}$ to vector bundles over
$\mathcal{M}$ one can construct the direct sum
$\oplus_{i=1}^n\func{V}_i$ and the tensor product
$\otimes_{i=1}^n\func{V}_i$.
\item Given a functor 
$\map{\func{V}}{\cat{Man}^n}{\cat{VBund}_{\mathbb{R}}}$ such that
$V\mathcal{M}$ is a vector bundle over $\mathcal{M}$, the canonical
pairing of $V\mathcal{M}$ and $V^*\mathcal{M}$ becomes a natural
transformation
$\nt{\langle,\rangle}{\func{V}^*\otimes\func{V}}{\func{\Lambda}^0}$
whose components cover the identity morphism.
\item For every functor
$\map{\func{V}}{\cat{Man}^n}{\cat{VBund}_{\mathbb{R}}'}$ and every
$r\in\mathbb{R}$ there is a natural transformation
$\nte{m_r}{\func{V}}{\func{V}}$ whose component at $\mathcal{M}$ is
given by the map $\map{m_r}{V\mathcal{M}}{V\mathcal{M}}$ such that
$m_r(v):=rv$. If $r\not=0$ this is a natural equivalence.
\item All of the functors above can be complexified, which
yields functors into $\cat{VBund}_{\mathbb{C}}$ or
$\cat{VBund}_{\mathbb{C}}'$. The
complexification of $\func{V}$ will be denoted by
$\func{V}_{\mathbb{C}}$ and there is a natural equivalence
$\nte{^-}{\func{V}_{\mathbb{C}}}{\func{V}_{\mathbb{C}}}$ in
$\cat{VBund}_{\mathbb{R}}$ (or $\cat{VBund}_{\mathbb{R}}'$) which
sends each section to its complex conjugate.
\item The above constructions (dual, direct sum, tensor product)
and natural transformations (pairing, $m_r$) can also be applied
directly to complex vector bundles in a canonical (Hermitean) way.
\end{enumerate}

It will be convenient to consider distributions and integration in
a categorical setting too:
\begin{definition}
$\cat{TVec}$ is the category of topological vector spaces
with injective continuous linear maps as morphisms. The functor
$\map{\mathbb{C}}{\cat{Man}^n}{\cat{TVec}}$ is the constant functor
$\mathbb{C}$, i.e.\ it assigns to each object the one dimensional
space $\mathbb{C}$ and to each morphism the identity morphism.

The \emph{functor of test-sections} is the functor
$\map{\func{C}^{\infty}_0}{\cat{VBund}_{\mathbb{C}}'}{\cat{TVec}}$
which maps each complex vector bundle $\mathcal{V}$ to the space
$\Test_0(\mathcal{V})$ of compactly supported smooth sections of
$\mathcal{V}$ in the test-section topology.\footnote{For a precise
definition of the well-known topologies on test-sections and smooth
sections we refer to \cite{Dieudonne} Ch.\ 17.} A morphism $\nu$,
covering a morphism $\mu$, is mapped to the push-forward $\nu_*$
defined by $\nu_*(f)=\nu\circ f\circ\mu^{-1}$ on
$\mu(\mathcal{M}_1)$, extended by $0$ to all of $\mathcal{M}_2$.

The \emph{functor of smooth sections} is the contravariant functor
$\map{\func{C}^{\infty}}{\cat{VBund}_{\mathbb{C}}}{\cat{TVec}}$
which maps each complex vector bundle $\mathcal{V}$ to the space
$\Test(\mathcal{V})$ of smooth sections of $\mathcal{V}$ in the
usual topology. A morphism $\nu$, covering a morphism $\mu$, is
mapped to the pull-back $\nu^*$ defined by
$\nu^*(f)=\nu^{-1}\circ f\circ\mu$.

The \emph{functor of distributions} is the contravariant functor
$\map{\func{Distr}}{\cat{VBund}_{\mathbb{C}}'}{\cat{TVec}}$
which maps each complex vector bundle $\mathcal{V}$ to the space
$(\Test_0(\mathcal{V}))'$ of distributions on $\mathcal{V}$ with
the weak topology induced by $\Test_0(\mathcal{V})$. A morphism
$\nu$, covering a morphism $\mu$, is mapped to the pull-back
$\nu^*$ defined by $\nu^*u:=u\circ\nu_*$.
\end{definition}
We will not need compactly supported distributions, but they can be
defined as the functor dual to $\func{C}^{\infty}$. Notice that
objects which are not compactly supported, such as smooth sections
or distributions, behave contravariantly, whereas compactly supported
ones behave covariantly. Also note that the pull-back of a smooth
section can only be defined for morphisms that restrict to
isomorphisms of the fibers. The following constructions will be of
importance in section \ref{sec_LCDirac}:
\begin{enumerate}
\item[12.] There is a natural transformation
$\nt{\int}{\func{C}^{\infty}_0\circ|\func{\Lambda}^n|}{\mathbb{C}}$
which assigns to each $\omega\in\Test_0(|\Lambda^n\mathcal{M}|)$ the
integral $\int_{\mathcal{M}}\omega$.
\item[13.] Let
$\map{\func{f}}{\cat{VBund}_{\mathbb{C}}}{\cat{VBund}_{\mathbb{C}}'}$
be the forgetful functor. For any functor 
$\map{\func{V}}{\cat{Man}^n}{\cat{VBund}_{\mathbb{C}}}$ there is a
canonical natural transformation
$\nt{\kappa}{\func{C}^{\infty}_0\circ\func{f}\circ\func{V}}
{\func{C}^{\infty}\circ\func{V}}$, whose components are the canonical
injections $\Test_0(V\mathcal{M})\subset\Test(V\mathcal{M})$.
\item[14.] For any functor
$\map{\func{V}}{\cat{Man}^n}{\cat{VBund}_{\mathbb{C}}}$ there is a
canonical natural transformation
$\nt{\iota}{\func{C}^{\infty}\circ(\func{V}\otimes|\func{\Lambda}^n|)}
{\func{Distr}\circ\func{f}\circ\func{V}^*}$ given by $\iota_{\mathcal{M}}
(f\otimes\omega):=\int_{\mathcal{M}}\langle .,f\rangle\ \omega$
for any smooth section $f$ of $V\mathcal{M}$ and any density $\omega$
on $\mathcal{M}$. Each component of $\iota$ is injective.
\item[15.] Given a pair of functors
$\map{\func{V}_i}{\cat{Man}^n}{\cat{VBund}_{\mathbb{C}}'}$, $i=1,2$,
any natural transformation $\nt{t}{\func{V}_1}{\func{V}_2}$ lifts to a
corresponding natural transformation
$\nt{T}{\func{C}^{\infty}_0\circ\func{V}_1}{\func{C}^{\infty}_0
\circ\func{V}_2}$ defined pointwise by
$T_{\mathcal{M}}f:=t_{\mathcal{M}}\circ f$. For
$\map{\func{V}_i}{\cat{Man}^n}{\cat{VBund}_{\mathbb{C}}}$ and a
natural transformation $\nt{t}{\func{V}_1}{\func{V}_2}$ we can
similarly define $\nt{T}{\func{C}^{\infty}\circ\func{V}_1}
{\func{C}^{\infty}\circ\func{V}_2}$ pointwise by
$T_{\mathcal{M}}f:=t_{\mathcal{M}}\circ f$.
\end{enumerate}
Where convenient we will often identify a functor
$\map{\func{V}}{\cat{Man}^n}{\cat{VBund}_{\mathbb{C}}}$ with the
functor $\func{f}\circ\func{V}$, omitting the forgetful functor, as this
rarely leads to confusion.

Next we add the structure of a semi-Riemannian metric:
\begin{definition}
The category $\cat{SRMan}^n$ of semi-Riemannian manifolds is the
subcategory of $\cat{Man}^n$ whose objects $M=(\mathcal{M},g)$ are
$\Test$ manifolds $\mathcal{M}$ of dimension $n$ with a semi-Riemannian
metric $g$ and whose morphisms $\map{m}{M_1}{M_2}$ are given by the
isometric morphisms in $\cat{Man}^n$, i.e.\ morphisms
$\map{\mu}{\mathcal{M}_1}{\mathcal{M}_2}$ such that
$\mu_*g_1=g_2|_{\mu(\mathcal{M}_1)}$.
\end{definition}
The extra structure gives rise to extra functors and natural
equivalences that are of interest to us:
\begin{enumerate}
\item[16.] The forgetful functor $\map{\func{f}}{\cat{SRMan}^n}{\cat{Man}^n}$
assigns to each $M=(\mathcal{M},g)$ the underlying manifold
$\mathcal{M}$ and to each morphism $m$ the underlying morphism
$\mu$ in $\cat{Man}^n$.
\item[17.] We will write $\func{T}$, respectively $\func{T}^*$, for the
functors $\func{T}\circ\func{f}$, respectively $\func{T}^*\circ\func{f}$,
from $\cat{SRMan}^n$ to $\cat{VBund}_{\mathbb{R}}$. There is then a
natural equivalence $\nte{G}{\func{T}}{\func{T}^*}$ whose component at
$M=(\mathcal{M},g)$ is given by the map
$\map{G_{\mathcal{M}}}{T\mathcal{M}}{T^*\mathcal{M}}$ such that
$v\mapsto g(v,.)$.
\item[18.] The functor
$\map{\func{F}}{\cat{SRMan}^n}{\cat{VBund}_{\mathbb{R}}}$ assigns to
each object $M$ the frame bundle $F\mathcal{M}$, i.e.\ the bundle
whose fiber at a point $x\in\mathcal{M}$ consists of all orthonormal
bases of $T_x\mathcal{M}$ in the metric $g$. This fiber is a subset
of $T^{\otimes n}\mathcal{M}$. A morphism $m$ is mapped to the
push-forward $\mu_*$ acting on
$F\mathcal{M}\subset T^{\otimes n}\mathcal{M}$.
\item[19.] The functor
$\map{\func{Cl}}{\cat{SRMan}^n}{\cat{VBund}_{\mathbb{R}}}$ assigns to
each object $M=(\mathcal{M},g)$ the Clifford bundle $Cl\mathcal{M}$,
which is the vector bundle whose fiber at $x\in\mathcal{M}$ is the
Clifford algebra of $(T_x\mathcal{M},g)$ viewed as a linear space. A
morphism $m$ is mapped to the push-forward acting on
$Cl\mathcal{M}\subset \oplus_{k=0}^nT^{\otimes k}\mathcal{M}$.
Note that $\func{Cl}$ is naturally equivalent to
$\func{\Lambda}\circ\func{f}$, because we ignore the algebraic
structure on these vector bundles.
\item[20.] We define the \emph{volume form functor}
$\map{\func{vol}}{\cat{SRMan}^n}{\cat{VBund}_{\mathbb{R}}}$ as
$\func{vol}:=|\func{\Lambda}^n|\circ\func{f}$. When
$\map{m}{M_1}{M_2}$ is a morphism and
$d\mathrm{vol}_i:=\sqrt{|\det g_i|}$ the metric induced volume form
on $M_i$, then $\func{vol}$ maps $d\mathrm{vol}_1$ to the restriction
of $d\mathrm{vol}_2$ to $m(\mathcal{M}_1)$. There is a
canonical natural equivalence from $\func{\Lambda}^0$ to $\func{vol}$,
which consists of multiplication with the metric induced volume form.
\item[21.] Similarly there are natural equivalences between any functor
$\map{\func{V}}{\cat{SRMan}^n}{\cat{VBund}_{\mathbb{C}}}$ and
$\func{V}\otimes|\func{\Lambda}^n|$. Therefore we obtain a
canonical natural transformation
$\nt{\iota}{\func{C}^{\infty}\circ\func{V}}
{\func{Distr}\circ\func{V}^*}$, the components of which are injective.
\end{enumerate}

\section{The classical Dirac field}\label{sec_CLDirac}

After these mathematical preliminaries we are now ready to start
constructing the classical free Dirac field (as a locally covariant
classical field). We will first describe the geometric and algebraic
constructions, before we discuss the Dirac equation and its
fundamental solutions. We close by investigating to what extent the
relations between the Dirac operator, charge conjugation and adjoint
map fix the structure of the theory and find that the non-uniqueness
can be characterised in terms of the cohomology of the category of
spin spacetimes.

\subsection{Geometric aspects}\label{ssec_Dgeom}

In order to describe the Dirac field we need to introduce the notion
of a spin structure on a spacetime, combining the geometric and the
algebraic results of section \ref{sec_math}. This is the purpose of
the current subsection.

The systems that we will consider are intended to model Dirac quantum
fields living in a (region of) spacetime which is endowed with a fixed
Lorentzian metric (a background gravitational field). Mathematically
these regions are modelled as follows:
\begin{definition}\label{def_man}
By the term \emph{globally hyperbolic spacetime} we will mean a
connected, Hausdorff, paracompact, $\Test$ Lorentzian manifold
$M=(\mathcal{M},g)$ of dimension $d=4$, which is oriented, time-oriented
and admits a Cauchy surface.

A subset $O\subset\mathcal{M}$ of a globally hyperbolic spacetime $M$ is
called \emph{causally convex} iff for all $x,y\in O$ all causal curves
in $\mathcal{M}$ from $x$ to $y$ lie entirely in $O$.

The category $\cat{Spac}$ is the subcategory of $\cat{SRMan}^n$ whose
objects are all globally hyperbolic spacetimes $M=(\mathcal{M},g)$ and
whose morphisms are isometric embeddings $\psi$ that preserve the
orientation and time-orientation and such that $\psi(\mathcal{M}_1)$ is
causally convex.
\end{definition}
Most notations we use concerning the causal structure of spacetimes are
standard, cf.\ \cite{Wald}. The importance of causally convex sets is
that for any morphism $\psi$ the causal structure of $M_1$ coincides
with that of $\psi(M_1)$ inside $M_2$:
\[
\psi(J_{M_1}^{\pm}(x))=J_{M_2}^{\pm}(\psi(x))
\cap\psi(\mathcal{M}_1),\quad x\in\mathcal{M}_1.
\]
If $O\subset\mathcal{M}$ is a connected open causally convex set, then
$(O,g|_O)$ defines a globally hyperbolic spacetime in its own right. In
this case there is a canonical morphism $I_{M,O}:O\rightarrow M$ given
by the canonical embedding $\map{\iota}{O}{\mathcal{M}}$. We will often
drop $I_{M,O}$ and $\iota$ from the notation and simply write
$O\subset M$.

Notice that there is a forgetful functor
$\map{\func{f}}{\cat{Spac}}{\cat{SRMan}^n}$ and that we can define the
functor $\map{\func{F}^{\uparrow}_+}{\cat{Spac}}{\cat{Bund}}$ of
oriented, time-oriented orthonormal frames $F^{\uparrow}_+\mathcal{M}$
for the tangent bundle, in analogy to section \ref{ssec_geom}. This is
a principal $\mathcal{L}_+^{\uparrow}$-bundle over $M$, where the
special ortochronous Lorentz group $\mathcal{L}_+^{\uparrow}$ acts from
the right, i.e., given $e=(x,e_0,\ldots,e_3)\in F_+^{\uparrow}M$, where
$x\in\mathcal{M}$ and $e_a\in T_xM$ such that $g_x(e_a,e_b)=\eta_{ab}$
and $e_0$ is future pointing, the action of $\Lambda$ is defined by
$R_{\Lambda}e=e'=(x,e'_0,\ldots,e'_3)$ where $e'_a=e_b\Lambda^b_{\ a}$.
\begin{definition}\label{def_sman}
A \emph{spin structure} on $M$ is a pair $(SM,\pi)$, where $SM$ is a
principal $Spin^0_{1,3}$-bundle over $M$, the \emph{spin frame bundle},
with a right action $R_S$, $S\in Spin^0_{1,3}$, and $\map{\pi}{SM}{FM}$,
the \emph{spin frame projection}, is a base-point preserving bundle
homomorphism such that
\[
\pi\circ R_S=R_{\Lambda(S)}\circ\pi,
\]
where $S\mapsto\Lambda(S)$ is the universal covering map (cf.\ Proposition
\ref{covering}).

A globally hyperbolic \emph{spin spacetime} $SM=(\mathcal{M},g,SM,\pi)$
is an object $M=(\mathcal{M},g)$ of $\cat{Spac}$ which is endowed with
the spin structure $(SM,\pi)$.

The category $\cat{SSpac}$ is the subcategory of $\cat{Bund}$ whose
objects are all globally hyperbolic spin spacetimes
$SM=(\mathcal{M},g,SM,\pi)$ and whose morphisms $\map{\chi}{SM_1}{SM_2}$
cover a morphism $\map{\psi}{M_1}{M_2}$ in $\cat{Spac}$ and satisfy
$\chi\circ(R_1)_S=(R_2)_S\circ\chi$ and $\pi_2\circ\chi=\psi_*\circ\pi_1$,
where $p_i$ are the bundle projections, $\pi_i$ the spin frame projections
and $\psi_*$ the push-forward.
\end{definition}
Note that a morphism acts as a diffeomorphism of the fibers, because it
intertwines the group action.

Every globally hyperbolic spacetime admits a spin structure, which need
not be unique \cite{Geroch1,Geroch2,Dimock,Lawson+}. We will regard
distinct spin structures on the same underlying spacetime as distinct spin
spacetimes.\footnote{There exists another approach to spinors, which
considers on each spacetime the Clifford bundle. This Clifford bundle is
functorial in its dependence on the spacetime, but it does not generally
define a spin structure. Indeed, at each point one can identify the Spin
group inside the fiber of the Clifford bundle, but there may not be any
projection from these Spin groups onto the frame bundle that intertwines
the actions of the structure groups, the obstruction being a topological
twist. (Conversely, every spin structure can be seen as a topologically
twisted copy of the Spin groups in the Clifford bundle.) Nevertheless,
it appears to provide sufficient structure to describe all the relevant
physics in a functorial way. We refer to \cite{Tolksdorf} for more
information on this approach.} Spinor and cospinor fields are sections of
vector bundles associated to the spin frame bundle. We will require that
the assignment of these vector bundles is functorial:
\begin{definition}\label{def_lcqf}
A \emph{locally covariant spinor bundle} is a functor
$\map{\func{V}}{\cat{SSpac}}{\cat{VBund}_{\mathbb{C}}}$, written as
$SM\mapsto V_{SM}$, $\chi\mapsto\nu$, such that $\chi$ and $\nu$ cover
the same morphism $\psi$ in $\cat{Spac}$ and such that each $V_{SM}$ is
a vector bundle associated to the spin frame bundle $SM$ through some
representation. The dual functor $\func{V}^*$ is called a \emph{locally
covariant cospinor bundle}. Smooth sections of $V_{SM}$, respectively
$V^*_{SM}$, are called (Dirac) \emph{spinors} (or spinor fields),
respectively \emph{cospinors} (cospinor fields).
\end{definition}
The condition in the definition of a locally covariant spinor bundle ensures
that the vector bundle $V_{SM}$ and the spin frame bundle $SM$ are both
bundles over the same spacetime $M$.

For definiteness we pick out the following standard choice of locally
covariant spinor and cospinor bundles:
\begin{definition}
The \emph{standard locally covariant Dirac spinor bundle}
$\map{\func{D}_0}{\cat{SSpac}}{\cat{VBund}_{\mathbb{C}}}$ is the locally
covariant spinor bundle which associates to each object $SM$ of
$\cat{SSpac}$ the associated vector bundle 
$D_0M=SM\times_{Spin^0_{1,3}}\mathbb{C}^4$ of $SM$ with the representation
$\pi_0$, and which maps each morphism $\map{\chi}{SM_1}{SM_2}$ to the
morphism $\map{\xi}{D_0M_1}{D_0M_2}$ given by
$\xi(\left[E,z\right]):=\left[\chi(E),z\right]$.

The \emph{standard locally covariant Dirac cospinor bundle} $\func{D}_0^*$
is the dual functor of $\func{D}_0$.
\end{definition}
Recall that a point in $D_0M$ consists of an equivalence class of pairs
$(E,z)\in SM\times\mathbb{C}^4$, where the equivalence is given by
\[
\left[R_SE,z\right]=\left[E,\pi_0(S)z\right].
\]
The dual functor $\func{D}_0^*$ then assigns to each $SM$ the dual vector
bundle $D_0^*M$ whose points are equivalence classes of pairs
$(E,w^*)\in SM\times(\mathbb{C}^4)^*$, where the equivalence is given by
$\left[R_SE,w^*\right]=\left[E,w^*\pi_0(S^{-1})\right]$. (Here we consider
$w^*\in(\mathbb{C}^4)^*$ as a row vector, whereas $z\in\mathbb{C}^4$ is
treated as a column vector.)

For any object $SM$ the unique connection $\nabla_{SM}$ on $TM$ which is
compatible with the metric, $\nabla_{SM} g=0$, can be described by an
$l_+^{\uparrow}$-valued one-form $(\mathbf{\Omega}_{SM})^b_{\ a}$ on the
orthonormal frame bundle $F_+^{\uparrow}M$ (cf.\ \cite{Kobayashi+} Ch.2
Proposition 1.1), where $l_+^{\uparrow}$ is the Lie-algebra of
$\mathcal{L}_+^{\uparrow}$, which can be identified with the tangent space
of the fiber of $F_+^{\uparrow}M$ at any point. For every local section
$e$ of $F_+^{\uparrow}M$ the pull-back
$\omega^b_{\ a}:=e^*(\mathbf{\Omega}^b_{\ a})$ consists exactly of the
connection one-forms of $\nabla_{SM}$ expressed in the orthonormal frame
$e_a$. The one-form $(\mathbf{\Omega}_{SM})^b_{\ a}$ can be pulled back
by the spin frame projection $\pi$ and lifted to a $spin^0_{1,3}$-valued
one-form $\mathbf{\Sigma}_{SM}$ on $SM$:
\[
\mathbf{\Sigma}_{SM}:=
(d\Lambda)^{-1}\pi^*((\mathbf{\Omega}_{SM})^b_{\ a})=
\frac{1}{4}p^*((\mathbf{\Omega}_{SM})^b_{\ a})g_bg^a,
\]
where the last equality uses Proposition \ref{covering}. The one-form
$\mathbf{\Sigma}_{SM}$ determines a connection on the spin frame bundle
$SM$. For any associated vector bundle $DM$ we then find a connection, also
denoted by $\nabla_{SM}$, determined by the connection one-forms
$\mathbf{\sigma}:=E^*(\mathbf{\Sigma}_{SM})$ in a local section $E$ of
$SM$, as represented on $DM$ (we will give an explicit expression for
$\mathbf{\sigma}$ in equation (\ref{spinconnection}) below). The connection
can be viewed as a map
$\map{\nabla_{SM}}{\Test_0(D_0M)}{\Test_0(T^*M\otimes D_0M)}$, which is a
component of a natural transformation\footnote{Alternatively we could have
written the connection as a natural transformation from the 1-jet bundle
extension of $\func{D}_0$ to $\func{T}^*\otimes\func{D}_0$.}
$\nt{\nabla}{\func{C}^{\infty}_0\circ \func{D}_0}
{\func{C}^{\infty}_0\circ(\func{T}^*\otimes\func{D}_0)}$. The Leibniz rule
allows us to extended it to mixed spinor-tensors, using
e.g.\ $\nabla_a\langle v,u\rangle=\langle \nabla_av,u\rangle+
\langle v,\nabla_au\rangle$.

\subsection{Adjoints, charge conjugation and the Dirac operator,}\label{ssec_Dstruc}

We now define the adjoint and charge conjugation maps on spinors and cospinors.
These are special cases of the Fundamental Theorem \ref{fundamentaltheorem}, using
the complex conjugate and adjoint matrices\footnote{On a general representation
space of complex dimension four one can define many complex conjugations and
Hermitean inner products. In order to obtain the desired equalities
involving adjoint and charge conjugate spinors later on we need these
two operations to be compatible, i.e.\ 
$\langle\overline{v},\overline{w}\rangle=\overline{\langle v,w\rangle}$.
Without loss of generality we can then use the standard complex
conjugation and Hermitean inner product on $\mathbb{C}^4$.}
(cf.\ \cite{Good}).
\begin{theorem}\label{ACmatrix}
For any irreducible complex representation $\pi$ of the Dirac algebra $D$
there are matrices $A,C\in GL(4,\mathbb{C})$ such that
\begin{eqnarray}\label{assumption}
A=A^*,&\pi(g_a)^*=A\pi(g_a)A^{-1},&An> 0,\\
\overline{C}C=I,&-\overline{\pi(g_a)}=C\pi(g_a)C^{-1}&\nonumber
\end{eqnarray}
for all future pointing time-like vectors $n\in M_0\subset D$.
We have for all $S\in Spin^0_{1,3}$:
\[
A=-C^*A^TC,
\]
\[
\pi(S)^*A\pi(S)=A,\quad
\pi(S^{-1})C^{-1}\overline{\pi(S)}=C^{-1}.
\]
Moreover, if $A',C'\in M(4,\mathbb{C})$ have the properties stated above
for the irreducible complex representation $\pi'$ of $D$, then there is an
$L\in GL(4,\mathbb{C})$, unique up to a sign, such that $L^*A'L=A$,
$(\overline{L})^{-1}C'L=C$ and $\pi=L^{-1}\pi'L$ on $D$.
\end{theorem}
\begin{proof*}
To prove the existence of $A$ and $C$ in the representation $\pi_0$ we
may take $A=A_0:=\gamma_0$, $C=C_0:=\gamma_2$ and check the required
properties straightforwardly. Note for example that
\[
\gamma_0n^a\gamma_a=\left(\begin{array}{cc}
n^0I+n^i\sigma_i&0\\
0&n^0I-n^i\sigma_i\end{array}\right)> 0,
\]
because $\det(n^0I\pm n^i\sigma_i)=n^2>0$ and
$Tr(n^0I\pm n^i\sigma_i)=2n^0>0$. To prove the existence of $A$ and $C$
in a general irreducible complex representation $\pi$ one writes
$\gamma_a=K\pi(g_a)K^{-1}$ by Theorem \ref{fundamentaltheorem}
and verifies that $A=K^*A_0K$ and $C=\overline{K}^{-1}C_0K$ will do.

Given $A',C'$ satisfying equation (\ref{assumption}) for $\pi'$ we can
fix $K\in GL(4,\mathbb{C})$ such that $\pi'=K\pi K^{-1}$ on $D$ and the
desired matrix $L$ must be $L=zK$ for some $z\not=0$ by the Fundamental
Theorem \ref{fundamentaltheorem}. Now set $\tilde{A}:=K^*A'K$ and
$\tilde{C}:=(\overline{K})^{-1}C'K$ and note that $\tilde{A}$ and
$\tilde{C}$ satisfy (\ref{assumption}) for $\pi$. Because the sets of
matrices $\pi(g_a)^*$ and $-\overline{\pi(g_a)}$ both satisfy the
relations (\ref{Clifford}) we must have $aA=\tilde{A}$ and
$cC=\tilde{C}$ for some non-zero complex factors $a$ and $c$, again by the
Fundamental Theorem. Also, $|c|=1$ because $\overline{C}C=I$ and $a>0$
because $A=A^*$ and $A\pi(n)>0$ for future pointing time-like vectors.
Hence, $|z|^2=a$ and $z=c\overline{z}$, which fixes $z$ (and $L$) up to a
sign. This proves the last statement.

The equation $A=-C^*A^TC$ holds for $A_0,C_0$ and therefore also in
general. For a unit vector $u=u^ag_a$ we have $u^2=\pm I$ and hence
\[
\pi(u)^*A\pi(u)=u^au^b\pi(g_a)^*A\pi(g_b)=u^au^bA\pi(g_ag_b)=A\pi(u^2)
=\pm A.
\]
For $S\in Spin_{1,3}$ we must therefore have that $\pi(S)^*A\pi(S)=\pm A$,
by definition of the Spin group. For $S=I$ the sign is a plus, so by
continuity and connectedness we conclude that $\pi(S)^*A\pi(S)=A$ for all
$S\in Spin^0_{1,3}$. For $C$ we use the fact that
\[
\pi(u^{-1})C^{-1}\overline{\pi(u)}=-\pi(u)^{-1}\pi(u)C^{-1}=-C^{-1}
\]
and hence $\pi(S^{-1})C^{-1}\overline{\pi(S)}=C^{-1}$ for all
$S\in Spin_{1,3}$.
\end{proof*}
Note that $g_5\in Spin_{1,3}\setminus Spin^0_{1,3}$. Indeed, using
$\pi_0$ and $A_0=\gamma_0$ in Theorem \ref{ACmatrix} we see that
$\gamma_5^*A_0\gamma_5=-A_0$, so $g_5\in Spin_{1,3}$ by definition, but not
in $Spin^0_{1,3}$.

In the following theorem we use the fact that for any pair of natural
transformations $\nt{t,t'}{\cat{SSpac}}{\cat{VBund}_{\mathbb{C}}'}$ we can
define the sum $t+t'$ and the tensor product $t\otimes t' $ componentwise.
\begin{theorem}\label{thm_Dstruc}
The standard locally covariant Dirac spinor and cospinor bundles admit natural
($\mathbb{C}$-antilinear) equivalences
$\nte{^+}{\func{D}_0}{\func{D}_0^*}$, $\nte{^c}{\func{D}_0}{\func{D}_0}$,
$\nte{^c}{\func{D}_0^*}{\func{D}_0^*}$ in $\cat{VBund}_{\mathbb{R}}$ and a
natural transformation $\nt{\gamma}{\func{D}_0}{\func{T}^*\otimes\func{D}_0}$
in $\cat{VBund}_{\mathbb{C}}'$ such that all components cover the identity
morphism and the following equations hold both on spinors and cospinors
(i.e.\ we denote the inverses of $^+$ and $^c$ by the same symbol):
\begin{eqnarray}\label{eqn_Dstruc}
^+\circ ^+=1=^c\circ ^c&&^+\circ ^c=-1\circ ^c\circ ^+\nonumber\\
\langle,\rangle\circ S\circ (^+\otimes ^+)=&^-\circ\langle,\rangle&
=\langle,\rangle\circ (^c\otimes ^c)\nonumber\\
(1\otimes^+)\circ\gamma=\gamma^*\circ^+&&
(1\otimes^c)\circ\gamma=-1\circ\gamma\circ^c
\end{eqnarray}
\[
(1+S\otimes 1)\circ (1\otimes \gamma)\circ\gamma=(2\circ g)\otimes 1,
\]
\[
\nabla\circ\gamma=\gamma\circ\nabla,
\]
where
$\nte{S}{\func{D}_0\otimes\func{D}^*_0}{\func{D}^*_0\otimes\func{D}_0}$
and $\nte{S}{\func{T}^*\otimes\func{T}^*}{\func{T}^*\otimes\func{T}^*}$
swap the factors in the tensor product,
$\nt{g}{\func{\Lambda}^0}{\func{T}^*\otimes\func{T}^*}$ maps the
function $1$ to the metric $g$ and
$\nt{\gamma^*}{\func{D}^*_0}{\func{T}^*\otimes\func{D}^*_0}$ is  the
adjoint map of $\gamma$ under the canonical pairing $\langle,\rangle$.
Futhermore, for every object $SM$, every time-like future pointing tangent
vector $n\in TM$ and every $v\in D_0M$
we have $\langle n\otimes v^+,\gamma(v)\rangle\ge 0$.
\end{theorem}
The natural transformation $\gamma$ can also be seen as a natural
transformation $\func{T}\Rightarrow\mathrm{End}(\func{D}_0)$ or
$\func{T}\Rightarrow\mathrm{End}(\func{D}_0^*)$. Equations
(\ref{eqn_Dstruc}) simply give the usual computational rules for spinors
and cospinors in a functorial setting. Thus, for every $SM$ and every
$p\in D_0M$, $q\in D_0^*M$ we have:
\begin{eqnarray}\label{eqn_Dstruc2}
p^{++}=p=p^{cc}&&p^{c+}=-p^{+c}\nonumber\\
\langle p^+,q^+\rangle=&\overline{\langle q,p\rangle}&=\langle q^c,
p^c\rangle\nonumber\\
(\gamma_{\mu}p)^+=p^+\gamma_{\mu}&&
(\gamma_{\mu}p)^c=-\gamma_{\mu}p^c\nonumber
\end{eqnarray}
\[
\gamma_{\mu}\gamma_{\nu}+\gamma_{\nu}\gamma_{\mu}=2g_{\mu\nu}I,\quad
\nabla_a\gamma_b\equiv 0,
\]
where we have dropped the subscript $SM$ to lighten the notation.
\begin{proof*}
The canonical pairing $\nt{\langle,\rangle}{\func{D}_0^*\otimes\func{D}_0}
{\func{\Lambda}^0_{\mathbb{C}}}$ on $SM$ is given by
$\langle\left[E,w^*\right],\left[E,z\right]\rangle=\langle w,z\rangle$,
where the right-hand side is the standard Hermitean inner product on
$\mathbb{C}^4$. Note that this is well-defined, because we can always get
the same $E\in SM$ on the left-hand side by a suitable action of
$Spin^0_{1,3}$. The components of the natural equivalences $^+$ and $^c$
on each $SM$ are defined using the matrices $A_0$ and $C_0$ of Theorem
\ref{ACmatrix} and their properties:
\begin{eqnarray}
\left[E,z\right]^c:=\left[E,C_0^{-1}\overline{z}\right]&&
\left[E,w^*\right]^c:=\left[E,\overline{w}^*C_0\right]\nonumber\\
\left[E,z\right]^+:=\left[E,z^*A_0\right]&&\left[E,w^*\right]^+:=
\left[E,A_0^{-1}w\right].\nonumber
\end{eqnarray}
These are well-defined isomorphisms in $\cat{VBund}_{\mathbb{R}}$ and
they give rise to natural equivalences satisfying the first two lines
of equation (\ref{eqn_Dstruc}).

Now fix $E\in SM$, let $e_a$ be the orthonormal basis
$(e_0,\ldots,e_3)=\pi(E)$ of $T_{p(E)}M$, where $\map{\pi}{SM}{FM}$ is
the spin frame projection, and let $e^a$ be the dual basis of
$T^*_{p(E)}M$. On $SM$ we define the component of the natural
transformation $\gamma$ on $SM$ to be
\[
\gamma(\left[E,z\right]):=e^a\otimes\left[E,\gamma_az\right].
\]
This is well-defined, because a different section $E':=R_SE$ gives rise
to the frame $e'_a=e_b\Lambda^b_{\ }a(S)$ and the dual frame
$(e')^a=\Lambda^a_{\ b}(S^{-1})e^b$ and on the other hand
$\pi_0(S^{-1})\gamma_a\pi_0(S)=\gamma_b\Lambda^b_{\ a}(S^{-1})$ by
definition of $\Lambda$ (Proposition \ref{covering}). $\gamma$ is indeed
a morphism in $\cat{VBund}_{\C}'$ and gives rise to a natural
transformation. The third line of equation (\ref{eqn_Dstruc}) follows
again from the properties of $A$ and $C$ (see Theorem \ref{ACmatrix}):
\begin{eqnarray}
\gamma(\left[E,z\right]^c)&=&e^a\otimes
\left[E,\gamma_a C_0^{-1}\overline{z}\right]
=-e^a\otimes\left[E,C_0^{-1}\overline{\gamma_az}\right]
=-(\gamma(\left[E,z\right]))^c,\nonumber\\
\gamma^*(\left[E,z\right]^+)&=&e^a\otimes\left[E,z^*A_0\gamma_a\right]
=e^a\otimes\left[E,z^*\gamma_a^*A\right]
=(\gamma(\left[E,z\right]))^+\nonumber
\end{eqnarray}
and similarly on cospinors. Also,
\begin{eqnarray}
\nabla_b\gamma_a&=&\sigma_b\gamma_a-\gamma_a\sigma_b
-\Gamma^c_{\ ba}\gamma_c=\frac{1}{4}\Gamma^c_{\ bd}
(\gamma_c\gamma^d\gamma_a-\gamma_a\gamma_c\gamma^d)-
\Gamma^c_{\ ba}\gamma_c\nonumber\\
&=&\frac{1}{4}\Gamma^c_{\ bd}(\gamma_c\left\{\gamma^d,\gamma_a\right\}
-\left\{\gamma_a,\gamma_c\right\}\gamma^d-4\delta^d_a\gamma_c)
=\frac{-1}{2}\Gamma^c_{\ bd}(\delta^d_a\gamma_c+\eta_{ac}\gamma^d)
=0.\nonumber
\end{eqnarray}

Finally, for every object $SM$, every future pointing tangent vector
$n\in TM$ and every $v\in D_0M$ we have
$\langle n\otimes v^+,\gamma(v)\rangle=\langle v^+,An^a\gamma_a v\rangle\ge 0$
again by Theorem \ref{ACmatrix}.
\end{proof*}

In terms of the Christoffel symbols $\Gamma^{\rho}_{\ \mu\nu}$, the
frame $e^a_{\rho}$ and representing $g_a$ on $D_0M$ using the
$\mathrm{End}(D_0M)$-valued one-forms $\gamma$, the connection one-forms
of the spin connection can be expressed
as\footnote{Note the sign error in \cite{Dimock,Fewster+}.}
\begin{eqnarray}\label{spinconnection}
\sigma_b&:=&\frac{1}{4}\Gamma^a_{\ bc}\gamma_a\gamma^c,\\
\Gamma^a_{\ bc}&=&-e^{\rho}_c(e_b^{\sigma}\partial_{\sigma} e^a_{\rho})+e^a_{\rho}e_b^{\mu}e_c^{\nu}\Gamma^{\rho}_{\ \mu\nu}.\nonumber
\end{eqnarray}
The Dirac operator is defined on spinors and cospinors by
\[
\dirop_{SM}:=\gamma^a\nabla_a.
\]
This defines natural transformations
$\nt{\dirop}{\func{C}_0^{\infty}\circ\func{D}_0}{\func{C}_0^{\infty}\circ\func{D}_0}$,
respectively
$\nt{\dirop}{\func{C}^{\infty}_0\circ\func{D}^*_0}{\func{C}^{\infty}_0\circ\func{D}^*_0}$.
The intertwining relations of the adjoint and charge conjugation with the Dirac
operator follow from their intertwining with $\gamma$ in Theorem
\ref{thm_Dstruc}:
\begin{proposition}\label{prop_Dirop}
$\dirop\circ ^+=^+\circ\dirop$, $\dirop\circ{\ }^c=-1\circ{\ }^c\circ\dirop$.
\end{proposition}
\begin{proof*}
Recall that $^+$ and $^c$ can be defined pointwise on test-sections. Hence, on any object
$SM$
\begin{eqnarray}
(\dirop v)^c&=&((\partial_av-v\sigma_a)\gamma^a)^c
=(\partial_a\overline{v}-\overline{v\sigma_a})\overline{\gamma^a}C
\nonumber\\
&=&-(\partial(\overline{v}C)-\overline{v}C\sigma_a)\gamma^a
=-\dirop(\overline{v}C)=-\dirop v^c,\nonumber\\
(\dirop u)^+&=&(\gamma^a(\partial_au+\sigma_au))^+
=(\partial_au^*+u^*\sigma_a^*)(\gamma^a)^*A\nonumber\\
&=&(\partial_a(u^*A)-u^*A\sigma_a)\gamma^a=
\dirop(u^*A)=\dirop u^+,\nonumber
\end{eqnarray}
where the minus sign in the last line appears because the order of the
two factors of $\gamma$ in the expression for $\sigma_a$ needs to be
changed. It follows that $(\dirop v)^+=(\dirop v^{++})^+
=(\dirop v^+)^{++}=\dirop v^+$ and $(\dirop u)^c=(\dirop u^+)^{+c}
=-(\dirop u^+)^{c+}=(\dirop u^{+c})^+=-(\dirop u^{c+})^+=-\dirop u^c$.
\end{proof*}

\begin{remark}
A change in the sign convention, $\tilde{\eta}:=-\eta$, has no physical
consequences. In fact, this simply gives rise to $D\simeq Cl_{3,1}$ as
the Dirac algebra, but since $Cl^0_{3,1}=Cl^0_{1,3}$ nothing changes in
the representation\footnote{Notice that a complex irreducible
representation of $Cl_{1,3}$ extends to an irreducible representation of
$M(4,\mathbb{C})$ and therefore also gives a complex irreducible
representation of $Cl_{3,1}$ and vice versa. The standard Clifford
algebra isomorphism $Cl_{3,1}\simeq M(4,\mathbb{R})$ appears if and only
if the representation of $Cl_{1,3}$ is a Majorana representation,
i.e.\ if $\overline{\gamma}_a=-\gamma_a$. In that case we also find
(see e.g.\ \cite{Choquet+} p.332)
\[
Pin_{3,1}\simeq \left\{S\in M(4,\mathbb{R})| \det S=1,
\forall v\in M_0 SvS^{-1}\in M_0\right\}\not=Pin_{1,3}.
\]}
of the group $Spin^0_{1,3}=Spin^0_{3,1}$. To accommodate this change one
can set $\tilde{\gamma}_a:=i\gamma_a$ in equation (\ref{Clifford}),
which yields the same Dirac algebra and other constructions (although
we do get signs for all covectors when raising or lowering indices with
$\tilde{\eta}$). This also implies that one should drop the factor $i$
in front of the Dirac operator in the Dirac equation (\ref{eqn_Dirac})
below, which ensures that $P_cP=PP_c$ will still be a wave operator. We
can also keep the same matrices $A,C$, which now must satisfy the
relations:
\[
-\tilde{\gamma}_a^*=A\tilde{\gamma}_aA^{-1},\quad
\overline{\tilde{\gamma}}_a=C\tilde{\gamma}_aC^{-1}.
\]
The spinor and cospinor bundle and the adjoint and charge conjugation
maps then remain the same and all the relations between these
operations and the Dirac operator remain valid.
\end{remark}

\subsection{The Dirac equation and its fundamental solutions}\label{ssec_Deqn}

The Dirac equation on spinor and cospinor fields, respectively, on a spin
spacetime $SM$ is
\begin{equation}\label{eqn_Dirac}
(-i\dirop+m)u=0\quad (i\dirop+m)v=0,
\end{equation}
where the constant $m\ge 0$ is to be interpreted as the mass of the field.
These equations can be derived as the Euler-Lagrange equations from the action
$\mathcal{S}_D:=\int \mathcal{L}_D$
with the Lagrangian density\footnote{The Lagrangian is a natural transformation
between the functor $\func{J_1D}_0$, which assigns to each spin spacetime $SM$ the
first-order jet bundle $J_1D_0M$ of the spinor bundle $D_0M$, to the functor
$\func{|\Lambda^n|}$ of densities. A component of this natural transformation
covers the identity morphism of $SM$ and is only a moprhism in $\cat{Bund}$, not
in $\cat{VBund}_{\R}'$, because it is not linear.}
\begin{equation}\label{eqn_Lagrangian}
\mathcal{L}_D:=\langle u^+,(-i\dirop+m)u\rangle d\mathrm{vol}_g
\end{equation}
by varying with respect to $u$ and $u^+$, viewed as independent fields. The
canonical momentum of the field $u$ on a Cauchy surface $C$ with future pointing
normal vector field $n$ is defined as
\begin{equation}\label{eqn_momentum}
\pi(x):=\frac{1}{\sqrt{-\mathrm{det}\ g(x)}}\frac{\delta \mathcal{S}_D}{\delta (n^{\mu}\nabla_{\mu}\psi(x))}
=-i\psi^+(x)\fsl{n}(x).
\end{equation}

We will write $P:=-i\dirop+m$ for the operator on spinors and $P_c:=i\dirop+m$
for the operator on cospinors. These are components of natural transformations
$\nt{P}{\func{C}^{\infty}_0\circ\func{D}_0}{\func{C}^{\infty}_0\circ\func{D}_0}$,
$\nt{P}{\func{C}^{\infty}\circ\func{D}_0}{\func{C}^{\infty}\circ\func{D}_0}$ and
$\nt{P_c}{\func{C}^{\infty}_0\circ\func{D}_0^*}{\func{C}^{\infty}_0\circ\func{D}_0^*}$,
$\nt{P_c}{\func{C}^{\infty}\circ\func{D}_0^*}{\func{C}^{\infty}\circ\func{D}_0^*}$,
which we denote by the same symbol. We then have by Proposition \ref{prop_Dirop}:
\begin{eqnarray}\label{eqn_Pprop}
P\circ{\ }^c={\ }^c\circ P&\quad& P_c\circ{\ }^c={\ }^c\circ P_c,\nonumber\\
P_c\circ ^+=^+\circ P&\quad& P\circ ^+=^+\circ P_c,
\end{eqnarray}
i.e.\ if a spinor field $u$ is a solution to the Dirac equation, then so are
$u^+$ and $u^c$. (The adjoint and charge conjugation of $u$ are defined
pointwise.)

For a distribution $v$ on $D_0M$ we define the transpose $P^*$ by
$\langle P^*v,u\rangle:=\langle v,Pu\rangle$ and similarly for $P_c$. In this
way the transposes give rise to natural transformations
$\nt{P^*}{\func{Distr}\circ\func{D}_0}{\func{Distr}\circ\func{D}_0}$ and
$\nt{P^*_c}{\func{Distr}\circ\func{D}^*_0}{\func{Distr}\circ\func{D}^*_0}$.
\begin{lemma}
Let $\nt{\iota}{\func{C}^{\infty}\circ\func{D}_0^*}{\func{Distr}\circ\func{D}_0}$
and $\nt{\iota}{\func{C}^{\infty}\circ\func{D}_0}{\func{Distr}\circ\func{D}^*_0}$
be the canonical natural transformations (see section \ref{ssec_geom} item
14). Then $P^*\circ\iota=\iota\circ P_c$ and $P_c^*\circ\iota=\iota\circ P$.
\end{lemma}
\begin{proof*}
This follows from the fact that for each object $SM$
$\int_M \langle u,\dirop v\rangle d\mathrm{vol}_g=
-\int_M \langle\dirop u,v\rangle d\mathrm{vol}_g$ if at least one of
$u\in\Test(D_0M)$ and $v\in\Test(D_0^*M)$ is complactly supported. This
in turn follows from
$\langle\dirop v,u\rangle+\langle v,\dirop u\rangle=\nabla_a\langle v,\gamma^a u\rangle$
and Gauss' law.
\end{proof*}

One can find unique advanced and retarded fundamental solutions for the Dirac equation,
both for spinors and cospinors \cite{Lichnerowicz,Dimock}:
\begin{theorem}\label{thm_fundsln}
There are unique natural transformations
$\nt{S^{\pm}}{\func{C}^{\infty}_0\circ\func{D}_0}{\func{C}^{\infty}\circ\func{D}_0}$
and 
$\nt{S_c^{\pm}}{\func{C}^{\infty}_0\circ\func{D}_0^*}{\func{C}^{\infty}\circ\func{D}_0^*}$
such that $S^{\pm}\circ P=P\circ S^{\pm}=\kappa$,
$S_c^{\pm}\circ P_c=P_c\circ S_c^{\pm}=\kappa$ and such that for each $u\in\Test_0(D_0M)$,
$v\in\Test_0(D_0^*M)$ we have $\mathrm{supp}(S^{\pm}u)\subset J^{\pm}(\mathrm{supp}(u))$,
$\mathrm{supp}(S_c^{\pm}u)\subset J^{\pm}(\mathrm{supp}(u))$. Moreover,
\begin{eqnarray}
S^{\pm}\circ{\ }^c={\ }^c\circ S^{\pm}&\quad& S^{\pm}_c\circ{\ }^c={\ }^c\circ S^{\pm}_c,
\nonumber\\
S^{\pm}_c\circ ^+=^+\circ S^{\pm}&\quad& S^{\pm}\circ ^+=^+\circ S^{\pm}_c,\nonumber
\end{eqnarray}
\[
\int\circ \langle,\rangle\circ (1\otimes S^{\pm})=
\int\circ \langle,\rangle\circ (S_c^{\mp}\otimes 1).
\]
\end{theorem}
\begin{proof*}
The components of $S^{\pm}$ and $S_c^{\pm}$ are the advanced ($-$) and retarded ($+$)
fundamental solutions for $P$ and $P_c$, which are given by
$S^{\pm}:=(i\dirop+m)E^{\pm}$ and $S_c^{\pm}:=(-i\dirop+m)E^{\pm}$ respectively,
where $E^{\pm}$ are the unique advanced and retarded fundamental solutions for the
normally hyperbolic operator
$(i\dirop+m)(-i\dirop+m)=(-i\dirop+m)(i\dirop+m)=\dirop^2+m^2$. We refer to
\cite{Dimock} Theorem 2.1 for a detailed proof of the existence and uniqueness of
these operators (see also \cite{Baer+} for the existence and uniqueness of $E^{\pm}$).

The naturality of $S^{\pm}$ and $S_c^{\pm}$ follows from their uniqueness and the
naturality of $P$ and $P_c$. In detail: for every morphism
$\map{\chi}{SM_1}{SM_2}$ and every $f\in\Test_0(D_0M_1)$ the unique smooth solution to
$Pu=\chi_*f$ on $M_2$ with $\mathrm{supp}(u)\subset J^{\pm}(\mathrm{supp}(\chi_*f))$
pulls back to a solution $v:=\chi^*u$ of $Pv=f$ on $M_1$ with
$\mathrm{supp}(v)\subset J^{\pm}(\mathrm{supp}(f))$. By uniqueness we must then have
$u=S^{\pm}\chi_*f$ and $\chi^*u=S^{\pm}f$, i.e.\ $\chi^*\circ S^{\pm}\circ\chi_*=S^{\pm}$.
The same holds for cospinors. The commutation of $S^{\pm}$ and $S_c^{\pm}$ with charge
conjugation and adjoints follows from equation (\ref{eqn_Pprop}).

For arbitrary $u\in\Test_0(D_0M)$ and $v\in\Test_0(D_0^*M)$ we can find a
$\phi\in\Test_0(M)$ which is identically $1$ on the compact set
$\mathrm{supp}(S^{\pm}u)\cap \mathrm{supp}(S_c^{\mp}v)$. We then compute:
\begin{eqnarray}
\int_M\langle v,S^{\pm}u\rangle &=&
\int_M\langle P_cS_c^{\mp}v,\phi S^{\pm}u\rangle=
\int_M\langle S_c^{\mp}v,P\phi S^{\pm}u\rangle\nonumber\\
&=&\int_M\langle S_c^{\mp}v,\phi P S^{\pm}u\rangle=
\int_M\langle S_c^{\mp}v,u\rangle,\nonumber
\end{eqnarray}
which proves the last claim.
\end{proof*}
We define the advanced-minus-retarded fundamental solutions $S:=S^--S^+$
and $S_c:=S_c^--S_c^+$, which are natural transformations
$\nt{S}{\func{C}^{\infty}_0\circ\func{D}_0}{\func{C}^{\infty}\circ\func{D}_0}$
and 
$\nt{S_c}{\func{C}^{\infty}_0\circ\func{D}_0^*}{\func{C}^{\infty}\circ\func{D}_0^*}$
respectively.

\subsection{The non-uniqueness of the functorial Dirac structure}\label{ssec_uniqueness}

We have seen that the (standard) structure of Dirac spinors and cospinors,
adjoints, charge conjugation and the Dirac operator is entirely determined by
the functor $D_0$ and the natural equivalences $^+$, $^c$ and $\gamma$. We
formalise this with a definition:
\begin{definition}
By a \emph{Dirac structure} $\mathcal{D}:=(\func{D},^+,^c,\gamma)$ we mean a
locally covariant spinor bundle $\func{D}$ with a dual bundle $\func{D}^*$,
natural equivalences $\nte{^+}{\func{D}}{\func{D}^*}$,
$\nte{^c}{\func{D}}{\func{D}}$, and $\nte{^c}{\func{D}^*}{\func{D}^*}$ in
$\cat{VBund}_{\R}$ and a natural transformation
$\nt{\gamma}{\func{D}}{\func{T}^*\otimes\func{D}}$ in $\cat{VBund}_{\C}'$,
all of whose components cover the identity morphism and satisfying the relations
(\ref{eqn_Dstruc}) and $\langle\gamma_{SM}(v^+,v),n\rangle\ge 0$ for every
time-like future pointing vector $n\in TM$.

We call $\mathcal{D}_0:=(\func{D}_0,^+,^c,\gamma)$ of Theorem \ref{thm_Dstruc}
the \emph{standard Dirac structure}.

The category $\cat{DStruc}$ has all Dirac structures as objects and its morphisms
$\map{t}{\mathcal{D}_1}{\mathcal{D}_2}$ are all natural transformations
$\nt{t}{\func{D}_1}{\func{D}_2}$ whose components are injective morphisms
covering the identity morphism and intertwining the adjoints, charge conjugation
and $\gamma$ as follows:
$$
^{+_2}\circ t=t\circ ^{+_1},\quad ^{c_2}\circ t=t\circ ^{c_1},\quad
\gamma_2\circ (t\otimes t)=\gamma_1.
$$
\end{definition}

For each Dirac structure one can perform the constructions of subsection
\ref{ssec_Deqn}.
Because the Dirac algebra $D$ has a unique irreducible complex representation
one might expect that the category $\cat{DStruc}$ admits a corresponding unique
initial object, perhaps up to isomorphism. This is an object from which there
exists a morphism into any other object. However, as we will explain in this
subsection there is a certain cohomological obstruction of the category
$\cat{SSpac}$ involved. We will first consider the standard Dirac structure,
which would be a good candidate for an initial object, and prove the following
weaker property:
\begin{proposition}\label{prop_uniqueness}
Any morphism $t$ from a Dirac structure $\mathcal{D}$ to the standard Dirac
structure $\mathcal{D}_0$ is an isomorphism.
\end{proposition}
\begin{proof*}
Let $\map{t}{\mathcal{D}}{\mathcal{D}_0}$ be a morphism. By the injectivity of
the components of $\nt{t}{\func{D}}{\func{D}_0}$ we see that the complex
dimension of the fiber of $DM$ is at most four. On the other hand, the vector
bundles $DM$ are modules for the Dirac algebra represented by $\gamma$.
Because this algebra is simple, and because equations (\ref{eqn_Dstruc})
exclude the trivial representation, we find that $DM$ must have complex
dimension at least four. Therefore, $\nt{t}{\func{D}}{\func{D}_0}$ must be a
natural equivalence and it follows that $\map{t}{\mathcal{D}}{\mathcal{D}_0}$
is an isomorphism.
\end{proof*}
\begin{corollary}\label{cor_Dstruciso}
If we construct a Dirac structure $\mathcal{D}_{\pi}$ analogous to
$\mathcal{D}_0$, but using a different representation $\pi$ and matrices $A,C$,
then $\mathcal{D}_{\pi}$ is isomorphic to $\mathcal{D}_0$.
\end{corollary}
\begin{proof*}
Because we use the same representation on all spacetimes we can construct a
natural equivalence $\nte{t}{\mathcal{D}_{\pi}}{\mathcal{D}_0}$ whose components are
of the form $t_{SM}(\left[E,z\right]):=\left[E,Lz\right]$ for some
$L\in GL(4,\mathbb{C})$ which is independent of $SM$ (cf.\ Theorem \ref{ACmatrix}).
\end{proof*}
\begin{corollary}\label{cor_uniqueD0}
If $\mathcal{D}:=(\func{D}_0,^{+_1},^{c_1},\gamma')$ is any Dirac structure with
the standard locally covariant Dirac spinor bundle $\func{D}_0$, then $\mathcal{D}$
is isomorphic to the standard Dirac structure $\mathcal{D}_0$.
\end{corollary}
\begin{proof*}
At each point $x$ in each object $SM$ we can view $\gamma'_a$ as matrices that
represent the Dirac algebra in a representation $\pi$. Using the Fundamental
Theorem \ref{fundamentaltheorem} we write $\gamma'_a=L\gamma_aL^{-1}$ for some
$L(x)\in GL(4,\C)$. As $\gamma'_a$ is well-defined on $\func{D}_0$ we must have
$\pi_0(S)\gamma'_a\pi_0(S^{-1})=\gamma'_b\Lambda^b_{\ a}(S)$ for all
$S\in\mathrm{Spin}^0_{1,3}$. This also holds for the matrices $\gamma$, so we
conclude from the Fundamental Theorem that $\pi_0(S)L(x)=c(x)L(x)\pi_0(S)$, where
$c\equiv 1$ bytaking $S=I$. We can now define a natural equivalence
$\nte{t}{\func{D}_0}{\func{D}_0}$ by $[E,z]\mapsto [E,L(p(E))z]$ such that
$\gamma'\circ t=t\circ \gamma$. If we also define
$^{+_2}:=t\circ^{+_1}\circ t^{-1}$ and $^{c_2}:=t\circ^{c_1}\circ t^{-1}$, then
$\mathcal{D}\Leftrightarrow (\func{D}_0,^{+_2},^{c_2},\gamma)\Leftrightarrow 
\mathcal{D}_0$, where the last equivalence follows from the previous Corollary.
\end{proof*}

In fact, the proof of Corollary \ref{cor_Dstruciso} shows that for any $SM$ the
quadruple $(DM,^+,^c,\gamma)$ is unique up to an isomorphism $t_{SM}$, if $DM$
has four dimensional complex fibers. The isomorphism $t_{SM}$ itself, however,
is only unique up to a sign. In other words, on each spin spacetime we find a
discrete $\Z_2$-symmetry that preserves all physical
relations.\footnote{This may be compared to \cite{Canarutto+}, who use complex
spinor structures and then find a local (gauge) symmetry instead of our more
restricted global symmetries.}

Consider two Dirac structures $\mathcal{D}$ and $\mathcal{D}'$ whose 
locally covariant spinor bundles $\func{D}$ and $\func{D}'$ have four
dimensional complex fibers. Comparing the action of these functors on
morphisms of $\cat{SSpac}$ one finds a diagram that commutes up to a sign.
The existence of an initial object in the category $\cat{DStruc}$ then
boils down to the question whether one can choose signs for all spin
spacetimes $SM$ in such a way that all the diagrams commute. The answer is
not at all obvious, but can be neatly formulated in terms of the first
Stiefel-Whitney class of the category $\cat{SSpac}$. To explain this we
will briefly recall the definition of cohomology groups for categories
(cf.\ \cite{Roberts+}).

If $\cat{C}$ is any category, we can first build a simplicial set from it
called the \emph{nerve} of the category (cf.\ \cite{Segal}). A 0-simplex is
simply an object of $\cat{C}$, a 1-simplex is a morphism between two objects,
a 2-simplex is a commutative triangle, etc. We will write $\Sigma_n$ for the
set of all $n$-simplices. For $n\ge 1$ every $n$-simplex has $n+1$
\emph{faces}, which are described by maps
$\map{\partial_j}{\Sigma_n}{\Sigma_{n-1}}$, $0\le j\le n$, which remove the
$j^{\mathrm{th}}$ vertex from the diagram.

To find the cohomology of $\cat{C}$ with values in an Abelian
group\footnote{\cite{Roberts+} also considers the non-Abelian case, which is
much more involved.} $G$, we define an \emph{$n$-cochain} with values in $G$
to be a map $\map{v}{\Sigma_n}{G}$. We denote the set of $n$-cochains with
values in $G$ by $C^n(G)$ and we define the \emph{coboundary} map
$\map{d}{C^n(G)}{C^{n+1}(G)}$ by
$$
dv(s):=\sum_{j=0}^{n+1}(-1)^jv(\partial_j s),\quad s\in\Sigma_{n+1},
$$
where we have written the group operation of $G$ additively. One checks that
$d^2=0$ and defines $v$ to be \emph{closed} iff $dv=0$ and \emph{exact} iff
$v=dt$ for some $(n-1)$-cochain $t$. The sets of closed and exact $n$-cochains
are denoted by $B^n(G)$ and $Z^n(G)$, respectively. They inherit an Abelian
group structure from $G$ and because $Z^n(G)\subset B^n(G)$ is necessarily
normal one can define the $j^{\mathrm{th}}$ cohomology group as the quotient
$H^n(G):=B^n(G)/Z^n(G)$.

Now let us return to the study of Dirac structures. Suppose that $\mathcal{D}$
and $\mathcal{D}'$ both have four dimensional complex fibers. Without loss of
generality we may assume that both Dirac structures coincide on each spin
spacetime, but the action of their locally covariant spinor bundles on a
morphism $\chi$ agrees only up to a sign $v(\chi)\in\left\{\pm 1\right\}$. We
can view $v:\chi\mapsto v(\chi)$ as a $1$-cochain on the category $\cat{SSpac}$
with values in $\Z_2=\left\{0,1\right\}$, where $0$ corresponds to $+1$ and
$1$ to $-1$). Notice that for a composition of morphisms
$\chi=\chi_1\circ\chi_2$ we find $v(\chi)=v(\chi_1)+v(\chi_2)$ in $\Z_2$,
because the Dirac structures are both functorial. In cohomological terms this
means precisely that $dv=0$.

If there is a natural equivalence $\nte{t}{\mathcal{D}}{\mathcal{D}'}$, then
the components $t_{SM}$ are automorphisms of the Dirac structure at each $SM$,
i.e.\ $t_{SM}=\pm 1$, that compensate for all the minus signs in $v$. If we
view $t$ as a $0$-cochain with values in $\Z_2$, this means exactly that
$v=dt$. So we have proved:
\begin{theorem}\label{thm_cohom}
The number of inequivalent Dirac structures whose locally covariant
spinor bundles have four dimensional complex fibers equals the number of
first Stiefel-Whitney classes of the category $\cat{SSpac}$, i.e.\ the
number of elements in $H^1(\Z_2)$.
\end{theorem}

\begin{remark}
For scalar and vector fields the problem above can be avoided in a
natural way. Taking $\mathcal{L}_+^{\uparrow}$ in the defining (four-vector)
representation, the vector bundle associated to $F_+^{\uparrow}M$ is just
the tangent bundle $TM$. A morphism in $\cat{Spac}$ determines a unique
morphism on the tangent bundle, so no topological obstructions occur.
Similarly for the scalar field, where one uses the trivial one dimensional
representation of $\mathcal{L}_+^{\uparrow}$, whose associated vector
bundle is $\Lambda^0(M)=M\times\mathbb{R}$. Again a morphism in
$\cat{Man}^n$ automatically determines a unique morphism on these
associated vector bundles, now by the requirement that the volume element
is preserved.

In general one is dealing with representations of $\mathrm{Spin}^0_{1,3}$
and associates to each morphism in $\cat{SSpac}$ an intertwining operator
between such representations. For the associated vector bundles of $SM$,
the physical requirements that we imposed on the bundle morphisms,
concerning the adjoint and charge conjugation maps and $\gamma$, reduce
the intertwiners exactly to a choice of lifting $\mathcal{L}_+^{\uparrow}$
to its double cover. In this way it leads to the same first
Stiefel-Whitney class that characterises the number of spin structures
on a manifold. For the general case it is expected that one needs a
non-Abelian cohomology theory to quantify the obstruction for finding
initial objects.
\end{remark}

\section{The locally covariant quantum Dirac field}\label{sec_LCDirac}

After our discussion of the classical Dirac field in section \ref{sec_CLDirac}
we now turn to the quantum Dirac field, its construction, its Hadamard states
and its relative Cauchy evolution.

\subsection{Quantisation of the free Dirac field}\label{ssec_quant}

First we will quantise the free Dirac field in a generally covariant way and
establish some of its properties. For this purpose we also present the main
ideas of locally covariant quantum field theory as introduced in
\cite{Brunetti+} (see also \cite{Fewster}).

In the following any quantum physical system will be described by a
topological $^*$-algebra $\alg{A}$ with a unit $I$, whose self-adjoint
elements are the observables of the system. An injective and continuous
$^*$-homomorphism expresses the notion of a subsystem, whereas a state
is desccribed by a normalised and positive continuous linear functional
$\omega$, i.e.\ $\omega(A^*A)\ge 0$ for all $A\in\alg{A}$ and
$\omega(I)=1$. The state space of $\alg{A}$ is the set of all states and is
denoted by $\alg{A}^{*+}_1$. Every state gives rise to a GNS-representation
$\pi_{\omega}$ (see \cite{Schmuedgen} Theorem 8.6.2.), which is
characterised uniquely, up to unitary equivalence, by the GNS-quadruple
$(\pi_{\omega},\mathcal{H}_{\omega},\Omega_{\omega},\mathscr{D}_{\omega})$.
Here $\mathcal{H}_{\omega}$ is the Hilbert space on which
$\pi_{\omega}(\alg{A})$ acts as (possibly unbounded) operators with the
dense, invariant domain
$\mathscr{D}_{\omega}:=\pi_{\omega}(\alg{A})\Omega_{\omega}$. The vector
$\Omega_{\omega}$ is cyclic and satisfies
$\omega(A)=\langle\Omega_{\omega},\pi_{\omega}(A)\Omega_{\omega}\rangle$
for all $A\in\alg{A}$. 

The collection of all systems forms a category $\cat{TAlg}$:
\begin{definition}\label{def_alg}
The category $\cat{TAlg}$ has as its objects all unital topological
$^*$-algebras $\alg{A}$ and as its morphisms all continuous and
injective $^*$-homomorphisms $\alpha$ such that $\alpha(I)=I$.

A \emph{locally covariant quantum field theory} is a (covariant)
functor $\map{\func{A}}{\cat{SSpac}}{\cat{TAlg}}$, written as
$SM\mapsto\alg{A}_{SM}$, $\chi\mapsto\alpha_{\chi}$.

A locally covariant quantum field theory $\func{A}$ is called
\emph{causal} if and only if any pair of morphisms
$\map{\psi_i}{SM_i}{SM}$, $i=1,2$, such that
$\psi_1(\mathcal{M}_1)\subset(\psi_2(\mathcal{M}_2))^{\perp}$ in
$\mathcal{M}$ yields $\left[\alpha_{\Psi_1}(\alg{A}_{SM_1}),
\alpha_{\Psi_2}(\alg{A}_{SM_2})\right]=\left\{0\right\}$ in
$\alg{A}_{SM}$.

A locally covariant quantum field theory $\func{A}$ satisfies the
\emph{time-slice axiom} iff for all morphisms $\map{\psi}{SM_1}{SM_2}$
such that $\psi(\mathcal{M}_1)$ contains a Cauchy surface for
$\mathcal{M}_2$ we have $\alpha_{\Psi}(\alg{A}_{SM_1})=\alg{A}_{SM_2}$.
\end{definition}

Notice that the condition
$\psi_1(\mathcal{M}_1)\subset(\psi_2(\mathcal{M}_2))^{\perp}$ is
symmetric in $i=1,2$. The causality condition formulates how the quantum
physical system interplays with the classical gravitational background
field, whereas the time-slice axiom expresses the existence of a causal
dynamical law.

We now fix a choice of Dirac structure
$\mathcal{D}:=(\func{D},^+,^c,\gamma)$, in order to turn the free Dirac
field into a locally covariant field theory. Because we want to impose
the canonical anti-commutation relations it will also be convenient to
quantise spinor and cospinor fields simultaneously by introducing the
following terminology:
\begin{definition}
The \emph{locally covariant double spinor bundle} is the covariant
functor $D\oplus D^*$. We define the following natural equivalences
and natural transformations on this bundle, indicated by their
components at $SM$:
\begin{eqnarray}
(p\oplus q)^c:=p^c\oplus q^c&& (p\oplus q)^+:=q^+\oplus p^+\nonumber\\
\gamma_{\mu}(p\oplus q):=(\gamma_{\mu}p)\oplus (\gamma_{\mu}q)
&&\langle p\oplus q,p'\oplus q'\rangle:=\langle p^+,p'\rangle
+\langle q',q^+\rangle\nonumber\\
R(p\oplus q):=p\oplus (-q).&&\nonumber
\end{eqnarray}

A \emph{double spinor (field)} is an element of $\Test(DM\oplus D^*M)$. A
\emph{double test-spinor (field)} is an element of $\Test_0(DM\oplus D^*M)$.
The adjoint, charge conjugation and other operations are defined pointwise.
We also define the operator $P:=P\oplus P_c$, its advanced ($-$) and
retarded ($+$) fundamental solutions $S^{\pm}(u\oplus v):=(S^{\pm}u)\oplus
(S_c^{\pm}v)$ and $S:=S^--S^+$.

The \emph{exterior tensor product} $\mathcal{V}_1\boxtimes \mathcal{V}_1$
of two vector bundles $\mathcal{V}_i$ with fiber $V_i$ over manifolds
$\mathcal{M}_i$, $i=1,2$, is the vector bundle over
$\mathcal{M}_1\times\mathcal{M}_2$ whose fiber is $V_1\otimes V_2$ and
whose local trivialisations are determined by
$(O_1\times O_2)\times(V_1\otimes V_2)$, where $O_i\times V_i$ are local
trivialisations of $\mathcal{V}_i$.

The \emph{Dirac Borchers-Uhlmann algebra} $\alg{F}^0_{SM}$ on a
spin spacetime $SM$ is the topological $^*$-algebra 
\[
\alg{F}^0_{SM}:=\oplus_{n=0}^{\infty}\Test_0((DM\oplus D^*M)^{\boxtimes n}),
\]
where the direct sum is algebraic (i.e.\ only finitely many non-zero summands
are allowed) and
\begin{enumerate}
\item the product is given by continuous linear extension of
$f_1\cdot f_2:=f_1\boxtimes f_2$,
\item the $^*$-operation is given by continuous antilinear extension of
\[
(f_1\boxtimes\ldots\boxtimes f_n)^*:=
f_n^+\boxtimes\ldots\boxtimes f_1^+,
\]
\item as a topological vector space $\alg{F}^0_{SM}$ is the strict inductive
limit $\alg{F}^0_{SM}=\cup_{N=0}^{\infty}\oplus_{n=0}^N
\Test_0((DM\oplus D^*M)^{\boxtimes n}|_{K_N^{\times n}})$, where $K_N$ is an
exhausting and increasing sequence of compact subsets of $\mathcal{M}$ and the
test-section space of the restricted vector bundle
$(DM\oplus D^*M)^{\boxtimes n}|_{K_N^{\times n}}$ is given the
test-section topology.
\end{enumerate}
\end{definition}
The topology of $\alg{F}^0_{SM}$ is such that a state is given by a sequence of
$n$-point distributional sections $\omega_n$ of $(DM\oplus D^*M)^{\boxtimes n}$.
A morphism $\map{\chi}{SM_1}{SM_2}$ in $\cat{SSpac}$ determines a unique
morphism $\map{\alpha_{\chi}}{\alg{F}^0_{SM_1}}{\alg{F}^0_{SM_2}}$ that is given
by the algebraic and continuous extension of the morphism
$DM_1\oplus D^*M_1\rightarrow DM_2\oplus D^*M_2$ that is supplied by the functor
$\func{D}$. Together with this map on morphisms the map $SM\mapsto\alg{F}^0_{SM}$
becomes a locally covariant quantum field theory
$\map{\func{F}^0}{\cat{SSpac}}{\cat{TAlg}}$. Our next task will be to divide out
the ideals that generate the dynamics and the canonical anti-commutation relations.

We define the natural transformation
$\nt{(,)}{(\func{C}^{\infty}_0\circ(\func{D}\oplus\func{D}^*))\otimes_{\mathbb{R}}
(\func{C}^{\infty}_0\circ(\func{D}\oplus\func{D}^*))}{\mathbb{C}}$ whose
components are the sesquilinear forms:
\[
(f_1,f_2):=i\int_M \langle f_1,RSf_2\rangle.
\]
Note that this is indeed a natural transformation, because it can be written as a
composition of natural transformations including $\int$, $\langle,\rangle$, $^+$ and
$\kappa$.
\begin{lemma}\label{lem_bracket}
On each object $SM$ the sesquilinear form $(,)$ is Hermitean,
$\overline{(f_1,f_2)}=(f_1^c,f_2^c)=(f_2,f_1)$, and there holds $(f_1^+,f_2^+)=(f_2,f_1)$.
For any spacelike Cauchy surface $C\subset M$ with future pointing unit normal vector
field $n^a$ we have
\begin{equation}\label{eqn_semidef}
(u_1\oplus v_1,u_2\oplus v_2)=\int_C \langle(Su_1)^+,\fsl{n}(Su_2)\rangle+
\langle S_cv_2,\fsl{n}(S_cv_1)^+\rangle.
\end{equation}
\end{lemma}
\begin{proof*}
The symmetry properties follow straightforwardly from the computational rules of
Theorems \ref{thm_Dstruc} and \ref{thm_fundsln}. For the last statement we also
need a partial integration (see e.g.\ \cite{Wald} equation (B.2.26) for Gauss' law)
and we use the Dirac equation:
\begin{eqnarray}
&&(u_1\oplus v_1,u_2\oplus v_2)\nonumber\\
&=&i\int_{J^+(C)}\langle P_cS_c^-u_1^+,Su_2\rangle+\langle P_cS_c^-v_2,Sv_1^+\rangle
+i\int_{J^-(C)}
\langle P_cS_c^+u_1^+,Su_2\rangle+\langle P_cS_c^+v_2,Sv_1^+\rangle\nonumber\\
&=&-\int_{J^+(C)}\nabla_a\langle S_c^-u_1^+,\gamma^aSu_2\rangle+
\nabla_a\langle S_c^-v_2,\gamma^aSv_1^+\rangle\nonumber\\
&&-\int_{J^-(C)}\nabla_a\langle S_c^+u_1^+,\gamma^aSu_2\rangle+
\nabla_a\langle S_c^+v_2,\gamma^aSv_1^+\rangle\nonumber\\
&=&\int_C n_a\langle S_c^-u_1^+,\gamma^aSu_2\rangle+
n_a\langle S_c^-v_2,\gamma^aSv_1^+\rangle
-\int_C n_a\langle S_c^+u_1^+,\gamma^aSu_2\rangle+
n_a\langle S_c^+v_2,\gamma^aSv_1^+\rangle\nonumber\\
&=&\int_C \langle(Su_1)^+,\fsl{n}(Su_2)\rangle+
\langle S_cv_2,\fsl{n}(S_cv_1)^+\rangle.\nonumber
\end{eqnarray}
\end{proof*}
From equation (\ref{eqn_semidef}) we notice that $(,)$ is positive semi-definite
and hence defines a (degenerate) inner product. We proceed by dividing
$\alg{F}^0_{SM}$ by the closed ideal $J_{SM}$ of $\alg{F}^0_{SM}$ generated by
all elements of the form $Pf$ or $f_1^+\cdot f_2+f_2\cdot f_1^+-(f_1,f_2)I$.
\begin{theorem}\label{thm_Ftheory}
The ideal $J_{SM}$ is a $^*$-ideal and for any morphism $\map{\chi}{SM_1}{SM_2}$
we have $\alpha_{\chi}(J_{SM_1})\subset J_{SM_2}$. We can define the locally
covariant quantum field theory $\map{\func{F}}{\cat{SSpac}}{\cat{TAlg}}$ which
assings to every spin spacetime $SM$ the $C^*$-algebra
$\alg{F}_{SM}:=\overline{\alg{F}^0_{SM}/J_{SM}}$.
\end{theorem}
\begin{proof*}
The elements that generate $J_{SM}$ are invariant under adjoints and under a
morphism they are mapped to elements of the same form. This proves the first
statement. It follows that the quotients $\alg{F}^0_{SM}/J_{SM}$ are topological
$^*$-algebras and that a morphism
$\map{\alpha_{\chi}}{\alg{F}^0_{SM_1}}{\alg{F}^0_{SM_1}}$ descends to the
quotients as a well-defined morphism. That each algebra $\alg{F}^0_{SM}/J_{SM}$
has a $C^*$-norm follows from the fact that they are the inductive limits
of finite dimensional Clifford algebras (\cite{Araki}). The morphisms on the
quotients are necessarily continuous in the norm and therefore extend to
morphisms on the $C^*$-algebras $\alg{F}_{SM}$.
\end{proof*}

\begin{definition}
A \emph{locally covariant quantum field in the locally covariant vector bundle}
$\func{V}$ for the locally covariant quantum field theory $\func{A}$ is a
natural transformation
$\nt{\Phi}{\func{C}_0^{\infty}\circ\func{V}^*}{\func{f}\circ\func{A}}$, where
we let $\map{\func{f}}{\cat{TAlg}}{\cat{TVec}}$ be the forgetful functor.

We define the locally covariant quantum fields
$\nt{B}{\func{D}\otimes \func{D}^*}{\func{F}}$,
$\nt{\psi}{\func{D}^*}{\func{F}}$ and
$\nt{\psi^+}{\func{D}}{\func{F}}$ by
$B_{SM}(f):=0\oplus f\oplus 0\oplus\ldots +J_{SM}$,
$\psi_{SM}(v):=B_{SM}(0\oplus v)$ and
$\psi^+_{SM}(u):=B_{SM}(u\oplus 0)$.
\end{definition}
That the latter really are locally covariant quantum fields is a consequence
of
\begin{proposition}\label{prop_psi}
The operator-valued maps $B_{SM},\psi_{SM},\psi^+_{SM}$ are $C^*$-algebra-valued
distributions and:
\begin{enumerate}
\item $P\circ\psi=0$ and $P_c\circ\psi^+=0$,
\item $\psi^+_{SM}(u)=\psi_{SM}(u^+)^*$,
\item $\left\{\psi_{SM}^+(u),\psi_{SM}(v)\right\}=(v^+\oplus 0,u\oplus 0)I
=-i\int_M \langle v,Su\rangle I$ and the other anti-commutators vanish.
\end{enumerate}
\end{proposition}
\begin{proof*}
The first item is $PB_{SM}(f)=B_{SM}(P^*f)=B_{SM}(Pf)=0$, where $P^*$ is the
formal adjoint of $P$. The last two items follow from the definitions of
$\psi_{SM}$ and $\psi^+_{SM}$ and the properties of $B_{SM}$ after a
straight-forward computation.

It remains to show that $\psi_{SM},\psi_{SM}^+$ are $C^*$-algebra-valued
distributions, because the result for $B_{SM}$ then follows. The $C^*$-subalgebra
of $\alg{F}_{SM}$ generated by $I,\psi_{SM}(v),\psi(v)_{SM}^*$ is a Clifford
algebra which is isomorphic to $M(2,\mathbb{C})$ and an explicit isomorphism is
given by $\psi_{SM}(v)\mapsto\left(\begin{array}{cc}0&\sqrt{c}\\ 0&0\end{array}
\right)$, where $c=(0\oplus v,0\oplus v)=-i\int_M \langle v,Sv^+\rangle>0$. It
follows that $\|\psi_{SM}(v)\|=\sqrt{c}$ is the operator norm of the
corresponding matrix, i.e.\footnote{The factor 2 in \cite{Fewster+}
Remark 2, p.340 seems to be erroneous.}\ 
\[
\|\psi_{SM}(v)\|^2=-i\int_M \langle v,Sv^+\rangle d\mathrm{vol}_g.
\]
In the test-spinor topology we then have continuous maps
$v\mapsto v\oplus v^+\mapsto -i\int_M \langle v,Sv^+\rangle$, from which
it follows that $v\mapsto\psi_{SM}(v)$ is norm continuous, i.e.\ it is
a $C^*$-algebra-valued distribution. The proof for $\psi_{SM}^+$
is analogous.
\end{proof*}
Note that the last two conditions of Proposition \ref{prop_psi} can also be
formulated in terms of natural transformations, because the algebraic
operations in $\alg{F}_{SM}$ can be expressed as such. The theory $\func{F}$
is the quantised free Dirac field and $\psi$ ($\psi^+$) is the locally
covariant Dirac (co)spinor field. Alternatively we could have used the
algebras $\alg{F}^0_{SM}/J_{SM}$ themselves instead of completing them to
$C^*$-algebras.

To see that the anti-commutator is the canonical one (cf.\ \cite{Lichnerowicz})
we apply Proposition 2.4c) of \cite{Dimock}, which says that
$S|_{C\times C}=-i\delta\fsl{n}$ for a Cauchy surface $C$ with future pointing
normal vector field $n$. Comparing with equation (\ref{eqn_momentum}) and using
$\fsl{n}^2=I$ we then find
\[
\left\{-i\psi^+_{SM}(\fsl{n}(x)),\psi_{SM}(y)\right\}=
-\int_M\langle y,S\fsl{n}x\rangle I=i\delta(y,x)I
\]
as expected.

So far our construction depends on the choice of a Dirac structure,
although naturally equivalent Dirac structures yield naturally equivalent
theories and quantum fields. The following theorem restricts attention to
the observable algebra, dividing out the freedom of choice completely and
yielding a unique theory, but for many purposes it is not convenient to
use it directly because it lacks locally covariant Dirac (co)spinor fields.
\begin{theorem}\label{thm_Btheory}
Let $\map{\func{B}}{\cat{SSpac}}{\cat{TAlg}}$ be the locally covariant
quantum field theory that assigns to each spin spacetime $SM$ the
$C^*$-subalgebra of $\alg{F}_{SM}$ generated by all even polynomials in
elements $B(f)$, with the induced action on morphisms. For all Dirac
structures with four dimensional complex fibers the resulting theories
$\func{B}$ are isomorphic.
\end{theorem}
\begin{proof*}
The algebras $\alg{B}_{SM}$ generated by the even polynomials are $C^*$-algebras.
Morphisms respect evenness and so restrict to morphisms on $\alg{B}$, making
$\func{B}$ a well-defined locally covariant quantum field theory. Now consider
two Dirac structures $\mathcal{D}$ and $\mathcal{D}_0$ with associated functors
$\func{F},\func{B}$ and $\func{F}_0,\func{B}_0$. If both Dirac structures have
four dimensional complex fibers, then we infer from the comment below Corollary
\ref{cor_Dstruciso} that there are $^*$-isomorphisms
$\map{\alpha_{SM}}{\alg{F}_{SM}}{(\alg{F}_0)_{SM}}$ such that for any
morphism $\map{\chi}{SM_1}{SM_2}$ we have
$\alpha_{SM_2}\circ\alpha_{\chi}=\epsilon_{\chi}\cdot (\alpha_0)_{\chi}
\circ\alpha_{SM_1}$, where $\epsilon_{\chi}=\pm 1$ depends only on $\chi$. It
follows from the evenness that the $\alpha_{SM}$ descend to $^*$-isomorphisms
$\map{\alpha_{SM}}{\alg{B}_{SM}}{(\alg{B}_0)_{SM}}$ that intertwine with the
morphisms. Hence, $\func{B}$ and $\func{B}_0$ are naturally equivalent.
\end{proof*}

\begin{proposition}\label{prop_LCQFTprop}
The locally covariant quantum field theory
$\map{\func{B}}{\cat{SSpac}}{\cat{TAlg}}$ of Theorem \ref{thm_Btheory} is
causal and satisfies the time-slice axiom.
\end{proposition}
\begin{proof*}
Causality follows from the anti-commutation relations,
\begin{eqnarray}
&&[B_{SM}(f_1)B_{SM}(f_2),B_{SM}(f_3)]\nonumber\\
&=&B_{SM}(f_1)\left\{B_{SM}(f_2),B_{SM}(f_3)\right\}-
\left\{B_{SM}(f_1),B_{SM}(f_3)\right\}B_{SM}(f_2)\nonumber\\
&=&(f_2,f_3)B_{SM}(f_1)-(f_1,f_3)B_{SM}(f_2),\nonumber
\end{eqnarray}
together with the support properties of $S$. For the time-slice axiom we
let $\map{\chi}{SM}{SM'}$ be a morphism in $\cat{SSpac}$, covering a
morphism $\map{\psi}{M}{M'}$ in $\cat{Spac}$, such that
$N:=\psi(M)\subset M'$ contains a Cauchy surface $C\subset M'$. Then we
can choose Cauchy surfaces $C^{\pm}\subset N$ such that
$C^{\pm}\subset I^{\pm}(C)$ and a smooth partition of unity
$\phi^+,\phi^-$ with $\mathrm{supp}\ \phi^{\pm}\subset J^{\pm}(C^{\mp})$.
Let $f\in\Test_0(DM\oplus D^*M)$ and write
\begin{equation}\label{timeslice1}
f=P(S^+f-\phi^+Sf)+\tilde{f},
\end{equation}
where $\tilde{f}:=P(\phi^+Sf)=-P(\phi^-Sf)$ is supported in
$J^+(C^-)\cap J^-(C^+)\subset N$ and $\phi^+Sf-S^+f$ has compact support.
Hence, $B_{SM_2}(f)=B_{SM_2}(\tilde{f})=\alpha_{\chi}(B_{SM_1}(\chi^*
(\tilde{f})))$. Because the algebra $\mathcal{F}_{M'}$ is generated by
such elements this shows that $\alpha_{\chi}$ is a $*$-isomorphism.
\end{proof*}

\begin{remark}
A Majorana spinor is a spinor $u$ such that $u=u^c$. In this case the
adjoint is anti-Majorana: $u^{+c}=-u^{c+}=-u^+$. We call a double spinor
$f=u\oplus v$ Majorana iff $u$ and $v^+$ are Majorana, which means that
$f^c=Rf$. Such spinors are sections of a subbundle of the Dirac spinor
bundle, which can be described by a Majorana representation. Notice that
every spinor is a unique complex linear combination of Majorana spinors.

To quantise Majorana spinors we note that
$\langle h^c,f\rangle=\langle h^+,f^{c+}\rangle$. This leads us to
define the charge conjugation on the quantised fields\footnote{Our
definition differs slightly from that of \cite{Dawson+}.} by
$\psi^c(v):=\psi^+(v^{c+})$ and $\psi^{+c}(u):=\psi(u^{c+})$, or
equivalently $B^c(f):=B(f^{c+})=B(f^c)^*$. We impose the Majorana
condition $B^c(f)=B(Rf)$ by dividing out the ideal generated by all
elements of the form $B(f-Rf^{c+})$. More precisely, if $\mathcal{H}$
is the Hilbert space obtained from $\Test_0(DM\oplus D^*M)$ by
dividing out the ideal of double spinors $f$ for which $(f,f)=0$, then
there is an orthogonal decomposition
$\mathcal{H}=\mathcal{H}_+\oplus\mathcal{H}_-$, where the elements in
$\mathcal{H}_{\pm}$ satisfy $Rf^{c+}=\pm f$. Indeed, every double
spinor can be written as $f=f_++if_-$, where
$f_{\pm}:=\frac{1}{2}(f\pm Rf^{c+})$ are in $\mathcal{H}_{\pm}$ and the
orthogonality follows from Lemma \ref{lem_bracket}. For the
$C^*$-algebraic quantisation we then have
$\mathcal{F}=\mathcal{F}_+\otimes \mathcal{F}_-$, where
$\mathcal{F}_-$ is the $C^*$-algebra of quantised Majorana spinors and
$\mathcal{F}_+$ the $C^*$-algebra of quantised anti-Majorana spinors
(see \cite{Bratteli+} section 5.2). The generators $\psi(v)$ and
$\psi^+(u)$ of $\mathcal{F}_-$ satisfy the additional relation
$\psi^c=\psi$ and $\psi^{+c}=-\psi^+$.
\end{remark}

\subsection{Hadamard states}\label{ssec_Had}

After Radzikowski's result \cite{Radzikowski} that a for a scalar field
state is of Hadamard form if and only if its wave front set has a certain
form, several people set out to extend this result to the Dirac field, or
more general quantum fields \cite{Kratzert,Hollands2,Sahlmann+}. All
three papers have provided an original contribution in their method of
proof, but upon careful analysis they all have minor gaps. We feel that
it is justified to comment on this here and to provide the necessary
results to fill any remaining gaps.

The most general results are the most recent ones, due to Sahlmann and
Verch \cite{Sahlmann+}, who set out to prove the equivalence of the
Hadamard form of a state, defined in terms of the Hadamard parametrix,
with a wave front set condition analogous to the scalar field case. One
of the techniques used is the scaling limit, but the proof of their
Proposition 2.8, which relates the wave front set of a distribution
to that of its scaling limit, is in our opinion insufficient (see the
footnote on page \pageref{ft_scaling}). In the appendix we prove a
similar statement as Proposition \ref{prop_scale}, thereby filling any
gap in \cite{Sahlmann+} and establishing the desired equivalence on a
firm ground. For the Dirac field, Hollands has proved that this wave
front set condition implies a specific form of the polarisation set
(\cite{Hollands} Theorem 4.1).

The scaling limit result can also be used to find the wave front sets of
the advanced and retarded fundamental solutions $E^{\pm}$ of normally
hpyerbolic operators on a globally hyperbolic spacetime, a result that
we prove as Theorem \ref{WFE}. Our proof is largely analogous to the work
of Radzikowski and the outcome is in direct analogy to the results of 
Duistermaat and H\"ormander \cite{Duistermaat+} for the scalar case. To
find the wave front sets of the fundamental solutions $S^{\pm}$ for the
Dirac equation we use (and correct) an idea of \cite{Hollands}.

Finally we comment on the results by Kratzert \cite{Kratzert}, which use a
spacetime deformation argument to compute the wave front set and polarisation
set of Hadamard states. This result has a gap, already identified in
\cite{Sahlmann+}, concerning the case of points $(x,\xi;y,\xi')$ where either
$\xi=0$ or $\xi'=0$, which prevents the propagation of the singularity from
the original to the deformed spacetime. This gap can be avoided using either
a propagation of Hadamard form result as in \cite{Sahlmann+}, or using the
commutation or anti-commutation relations and the explicit form of $WF(E)$,
respectively $WF(S)$. The latter argument, which appears to be implicit in
Radzikowski's paper \cite{Radzikowski}, works as follows: when
$(x,\xi;y,0)\in WF(\omega_2)$ then also $(y,0;x,\xi)\in WF(\omega_2)$ by the
(anti-)commutation relations and the fact that $WF(E)$ (or $WF(S)$) has no
points with either entry equal to $0$. Using the calculus of
Hilbert-space-valued distributions, Theorem \ref{WFHvalued}, we then find
that both
$(x,\xi;x,-\xi)\in WF(\omega_2)$ and $(x,-\xi;x,\xi)\in WF(\omega_2)$.
Because $\xi\not=0$ (by definition the wave front set does not contain the
zero covector) these points can both be propagated into a deformed spacetime,
where $WF(\omega)$ is known to satisfy the required microlocal condition.
This, however, leads to a contradiction, because
$WF(\omega_2)\cap -WF(\omega_2)=\emptyset$ and hence $\xi=0$. Therefore,
$WF(\omega_2)$ cannot contain points with one of the covectors equal to $0$.

After these historical notes we feel free to define the notion of Hadamard
states directly in terms of a wave front set condition, rather than using
the Hadamard parametrix. If $\omega$ is a state on $\alg{F}_{SM}$ then we
may consider the GNS-representation
$(\mathcal{H}_{\omega},\pi_{\omega},\Omega_{\omega})$ associated to $\omega$
and the $\mathcal{H}_{\omega}$-valued distribution on $DM\oplus D^*M$
defined by:
\[
v_{\omega}(f):=\pi_{\omega}(B_{SM}(f))\Omega_{\omega}.
\]
\begin{definition}
A state $\omega$ on $\alg{F}_{SM}$ is called \emph{Hadamard} if and
only if
\[
WF(v_{\omega})=\mathcal{N}^+:=\left\{(x,\xi)\in T^*M|\
\xi^2=0,\ \xi^{\mu}\mathrm{\ is\ future\ pointing\ or\ }0\right\}.
\]
A state $\omega$ on $\alg{B}_{SM}$ is called Hadamard if and only if
it can be extended to a Hadamard state on $\alg{F}_{SM}$. The set of
all Hadamard states on $\alg{B}_{SM}$ will be denoted by
$\mathcal{S}_{SM}$.
\end{definition}
Note that every state on $\alg{B}_{SM}$ can be extended to
$\alg{F}_{SM}$, by the Hahn-Banach Theorem and Proposition
\ref{prop_psi}. The Hadamard condition is independent of the
choice of extension, because it depends solely on the two-point
distribution as the following proposition shows (cf.\ \cite{Sahlmann+}, we
give a short proof using the more advanced microlocal techniques
developed in the appendix).

\begin{proposition}\label{prop_HadChar}
For a state $\omega$ on $\alg{F}_{SM}$ the following conditions are
equivalent:
\begin{enumerate}
\item $\omega$ is Hadamard,
\item $WF(v_{\omega})\subset\mathcal{N}^+$,
\item the two-point distribution
$\omega_2(f_1,f_2):=\omega(B_{SM}(f_1)B_{SM}(f_2))$ has
\[
WF(\omega_2)=\mathcal{C}:=\left\{(x,-\xi;y,\xi')\in
T^*M^{\times 2}\setminus\mathcal{Z}|\ (x,\xi)\sim (y,\xi'),
(x,\xi)\in\mathcal{N}^+\right\},
\]
where $(x,\xi)\sim (y,\xi')$ if and only if there is an
affinely parameterised light-like geodesic from $x$ to $y$ to which
$\xi,\xi'$ are cotangent,
\item there is a two-point distribution $w$ such that
$\omega_2(f_1,f_2)=iw(Pf_1,f_2)$ and $WF(w)=\mathcal{C}$.
\end{enumerate}
\end{proposition}
\begin{proof*}
First note that $\omega_2$ is a bidistribution on $DM\oplus D^*M$,
because $B_{SM}$ is an $\alg{F}_{SM}$-valued distribution and
multiplication in $\alg{F}_{SM}$ and $\omega$ are continuous. By Theorem
\ref{WFHvalued} the third statement implies the first, which trivially
implies the second. To show that the second statement implies the third
we use the argument of \cite{Strohmaier+}, Proposition 6.1. By Theorem
\ref{WFHvalued} we see that
$WF(\omega_2)\subset\mathcal{N}^-\times\mathcal{N}^+$, where
$\mathcal{N}^-:=-\mathcal{N}^+$. Defining
$\tilde{\omega}_2(f_1,f_2):=\omega_2(f_2,f_1)$ we find
$WF(\tilde{\omega}_2)\cap WF(\omega_2)=\emptyset$. Now,
$(\omega_2+\tilde{\omega}_2)(f_1,f_2)=i\int_M\langle f_1,RSf_2\rangle$, so
$WF(\omega_2)\cup WF(\tilde{\omega}_2)=WF(S)=WF(E)$ by Proposition
\ref{WFS} and hence
$WF(\omega_2)=WF(E)\cap\mathcal{N}^-\times\mathcal{N}^+=\mathcal{C}$
by Corollary \ref{cor_WFE}.

Now, assume that $\omega_2(f_1,f_2)=iw(Pf_1,f_2)$, where 
$WF(w)=\mathcal{C}$. Then
$WF(\omega_2)= WF((P^*\otimes I)w)\subset WF(w)=\mathcal{C}$. It follows
that $WF(v_{\omega})\subset\mathcal{N}^+$. For the converse we suppose that
$\omega$ is Hadamard and we choose a smooth real-valued function $\phi^+$
on $M$ such that $\phi^+\equiv 0$ to the past of some Cauchy surface $C_-$
and such that $\phi^-:=1-\phi^+\equiv 0$ to the future of another Cauchy
surface $C_+$. We then define
$w(f_1,f_2):=-i\omega_2(\phi^+S^-f_1+\phi^-S^+f_1,f_2)$. Note that
$w$ is a bidistribution which is well-defined, because $\phi^+S^-f_1$
and $\phi^-S^+f_1$ are compactly supported. By construction
$iw(Pf_1,f_2)=\omega_2(f_1,f_2)$. We now estimate the wave front set
of $w$ as follows. The wave front sets of $S^{\pm}$ are determined in
Proposition \ref{WFS}. Then we may apply Theorem 8.2.9 and 8.2.13 in
\cite{Hoermander} (in combination with equation (\ref{eqn_WFBvalued}))
to estimate the wave front sets of the tensor products
$\phi^{\pm}(x)S^{\mp}(x,y)\delta(x',y')$ and the compositions in
$iw(x,x')=\sum_{\pm}\int\omega_2(y,y')(\phi^{\pm}(x)S^{\mp}(x,y)
\delta(x',y'))$ respectively and, using
$WF(\omega_2)=\mathcal{C}$, we find:
\[
WF(iw)\subset \cup_{\pm} WF(S^{\mp}\otimes\delta)\circ
WF(\omega_2)\subset WF(\omega_2)=WF((P^*\otimes I)w)\subset WF(w),
\]
i.e.\ $WF(w)=WF(\omega_2)=\mathcal{C}$.
\end{proof*}

The second characterisation in Proposition \ref{prop_HadChar} is
especially useful, because it shows we do not need to compute the entire
wave front set, as long as we can estimate it. Employing similar
techniques as above one can use the anti-commutation relations and the
wave front set of $\omega_2$ to estimate the wave front sets of all
higher $n$-point distributions \cite{KS1}, showing that a Hadamard
state necessarily satisfies the microlocal spectrum condition ($\mu$SC)
of \cite{Brunetti+2} and it follows that the set of such states is closed
under operations from the algebra. We formulate this and other properties
of Hadamard states in the following

\begin{proposition}\label{prop_Hadprop}
The set $\mathcal{S}_{SM}$ of all Hadamard states on $\alg{B}_{SM}$ 
satisfies:
\begin{enumerate}
\item $\alpha_{\chi}^*(\mathcal{S}_{SM_1})\subset\mathcal{S}_{SM_2}$
for every morphism $\map{\chi}{SM_1}{SM_2}$,
\item $\mathcal{S}_{SM}$ is closed under operations from $\alg{B}_{SM}$,
\item $\alpha_{\chi}^*(\mathcal{S}_{SM_1})=\mathcal{S}_{SM_2}$
for every morphism $\map{\chi}{SM_1}{SM_2}$ such that
$\psi(\mathcal{M}_1)$ contains a Cauchy surface of $\mathcal{M}_2$.
\end{enumerate}
\end{proposition}
\begin{proof*}
The first property follows from Theorem \ref{prop_HadChar} and the fact
that wave front sets are local and geometric objects (cf.\ \cite{Hoermander}
Ch. 8). The second property relies on the anti-commutation relations, which
implies that the truncated $n$-point distributions are totally anti-symmetric
(cf.\ \cite{KS1,KS2}). The final property follows from the second
characterisation in Theorem \ref{prop_HadChar}, equation (\ref{eqn_WFBvalued})
in the appendix, the equation of motion and the Propagation of Singularities
Theorem for the wave front set, which in this case follows from the
propagation of the polarisation set \cite{Dencker}.
\end{proof*}

One can also prove that the state spaces are locally physically equivalent
\cite{Fewster} and that all quasi-free Hadamard states are locally
quasi-equivalent \cite{DAntoni+}. Whether the latter remains true for all
Hadamard states appears to be unknown.

We conclude this section with the remark that the functor
$\map{\func{S}}{\cat{SSpac}}{\cat{TVec}}$ defined by
$SM\mapsto\mathcal{S}_{SM}$ and $\chi\mapsto\alpha^*_{\chi}$ (restricted to
the relevant state space) is a locally covariant state space for the theory
$\func{B}$ \cite{Brunetti+}.

\subsection{The relative Cauchy evolution of the Dirac field and the
stress-energy-momentum-tensor}\label{ssec_RCE}

Now that we have a locally covariant free Dirac field at our
disposal we will investigate the idea of relative Cauchy evolution
for this field and prove that it yields commutators with the
stress-energy-momentum tensor. This result is completely analogous
to the result for the free scalar field of \cite{Brunetti+}.

Suppose that we have two objects $M_0=(\mathcal{M},g_0,SM_0,p_0)$
and $M_g=(\mathcal{M},g,SM_g,p_g)$ in $\mathfrak{SSpac}$, where
$\mathcal{M}$ is the same in both cases and such that outside a compact
set $K\subset\mathcal{M}$ we have $g=g_0$, $SM_g=SM_0$ and $p_g=p_0$. Now
let $N^{\pm}\subset M_0$ be causally convex open regions, each containing
a Cauchy surface for $M_0$, such that $K$ lies to the future of $N^-$
(i.e.\ $K\subset J^+(N^-)\setminus N^-$ in $M_0$ and hence also in $M_g$)
and to the past of $N^+$. We view $N^{\pm}$ as objects in
$\mathfrak{SSpac}$ and consider the canonical morphisms
$\map{\iota^{\pm}_0}{N^{\pm}}{M_0}$ and $\map{\iota^{\pm}_g}{N^{\pm}}{M_g}$.
By the time-slice axiom, Proposition \ref{prop_LCQFTprop}, these give
rise to $*$-isomorphisms
$\map{\beta^{\pm}_0}{\mathcal{B}_{N^{\pm}}}{\mathcal{B}_{M_0}}$ and
$\map{\beta^{\pm}_g}{\mathcal{B}_{N^{\pm}}}{\mathcal{B}_{M_g}}$. We
then define
\[
\beta_g:=\beta^+_0\circ(\beta^+_g)^{-1}\circ\beta^-_g\circ(\beta^-_0)^{-1}.
\]
The $*$-isomorphism $\map{\beta_g}{\alg{B}_{M_0}}{\mathcal{B}_{M_0}}$
measures the change in an operator $A\in\mathcal{B}_{N^-}$ as it evolves
to $N^+$ in the metric $g$ instead of $g_0$.\footnote{In \cite{Brunetti+}
it seems the authors have the scattering of a state in mind as it passes
through the perturbed metric, which leads them to consider the
$*$-isomorphisms $\beta_{g^{-1}}$ rather than $\beta_g$. When we take the
variation w.r.t.\ $g$ this gives rise to a sign.}\label{ft_RCEsign}
$\beta_g$ can be extended to a $*$-isomorphism of the algebra
$\mathcal{F}_{M_0}$, where we fix the signs for the isomorphisms
between the spinor bundles involved by identifying the double spinor
bundles over $N^{\pm}\subset M_0$ and $N^{\pm}\subset M_g$. It represents
the \emph{relative Cauchy evolution} of the free Dirac field.

We will want to compute the variation of the $*$-isomorphism $\beta_g$
as well as that of the action for the free Dirac field with respect to
the metric $g$. For this purpose we will suppose that the compact set
$K\subset\mathcal{M}$ has a contractible neighbourhood $O$ which does not
intersect either $N^{\pm}$. Let $\epsilon\mapsto g_{\epsilon}$ be a
smooth curve from $[0,1]$ into the space of Lorentzian metrics on
$\mathcal{M}$ starting at $g_0$ and such that $g_{\epsilon}=g_0$ outside
$K$ for every $\epsilon$. The spin bundle $SM_{\epsilon}$ must be trivial
over the contractible region $O$. If we assume it to be diffeomorphic to
$SM_0$ outside $K$ we can simply take $SM_{\epsilon}=SM_0$ as a manifold
and, choosing a fixed representation and matrices $A,C$, we obtain
$DM_{\epsilon}=DM$.

The deformation of the spin structure is contained entirely in the spin
frame projection $\map{\pi_{\epsilon}}{SM_0}{FM_{\epsilon}}$. Let $E$ be
a section of $SM_0$ over $O$ and set $(e_{\epsilon})_a:=\pi_{\epsilon}(E)$.
We require that $e_{\epsilon}$ varies smoothly with $\epsilon$ and that
$(e_{\epsilon})_a=(e_0)_a$ outside $K$. To show that projections
$\pi_{\epsilon}$ with these properties exist we can apply the Gram-Schmidt
orthonormalisation procedure to $(e_0)_a$ for all $\epsilon$
simultaneously. The assignment $E\mapsto e_{\epsilon}$ determines
$\pi_{\epsilon}$ completely, using the intertwining properties. The family
of frames $e_{\epsilon}$ determines principal fiber bundle isomorphisms
$FM_{\epsilon}\rightarrow FM_0$ between the frame bundles by
\[
\lambda_{\epsilon}:\left\{(e_\epsilon)_a\right\}\mapsto\left\{(e_0)_a\right\}
\]
on $K$ and extending it by the identity on the rest of $\mathcal{M}$.
By definition $f_{\epsilon}$ intertwines the action of
$\mathcal{L}_+^{\uparrow}$ on the orthonormal frame bundles.

\begin{remark}\label{rem_variation}
There may be many deformations of the spin structure, i.e.\ many families
of projections $\pi_{\epsilon}$ which satisfy our requirements. However,
the variation of terms like $\langle v,P_{\epsilon} u\rangle$ will not depend
on this choice. Indeed, if $\pi'_{\epsilon}$ is a different deformation of
the spin structure, then $e'_{\epsilon}:=\pi'_{\epsilon}(E)=
R_{\Lambda_{\epsilon}}e_{\epsilon}=\pi_{\epsilon}(R_{S_{\epsilon}}E)$
for some smooth curve $S_{\epsilon}$ in $Spin^0_{1,3}$. However, using the
invariance of $\langle,\rangle$ under the action of the gauge group
$Spin^0_{1,3}$, the variation will be equal in both cases. (Also
$\delta u=0$ for every spinor $u$, because $D_{\epsilon}M=DM$.) In this
sense the variation will only depend on the variation of the metric.
\end{remark}

\subsubsection{The stress-energy-momentum tensor}\label{ssec_SEMT}

The classical stress-energy-momentum tensor for the Dirac field is
defined as a variation of the action $\mathcal{S}=\int_{\mathcal{M}}\mathcal{L}_D$,
with the Lagrangian density (\ref{eqn_Lagrangian}), with respect to $g^{\mu\nu}(x)$:
\begin{equation}\label{eqn_SEMT1}
T_{\mu\nu}(x):=\frac{2}{\sqrt{-\mathrm{det}\ g(x)}}\frac{\delta S}{\delta g^{\mu\nu}(x)},
\end{equation}
where $\psi$ is a free classical Dirac spinor, $\psi^+$ its adjoint. An explicit
computation yields\footnote{For explicit computations we refer to section 4 of
\cite{Forger+}, who use a Lagrangian that differs from ours by a total derivative.
Varying with respect to $g_{\mu\nu}$ would yield the opposite sign.}
\[
T_{\mu\nu}=\frac{i}{2}\left(\langle\psi^+,\gamma_{(\mu}\nabla_{\nu)}\psi\rangle
-\langle\nabla_{(\mu}\psi^+,\gamma_{\nu)}\psi\rangle\right),
\]
Here the brackets around indices denote symmetrisation as an idempotent operation
and in the following indices between $|\ldots|$ are to be excluded from
the symmetrisation.

Following \cite{Fewster+} we quantise the stress-energy-momentum tensor via a
point-split procedure, i.e.\ we want to find a bi-distribution of scalar
test-functions which reduces to $T_{\mu\nu}$ on the diagonal and which can be
quantised in a straight-forward way. For this purpose we use a local spin frame
$E_A$ and recall that the components $\gamma^{\ A}_{a\ B}$ of $\gamma_a$ are
constant. We define:
\[
T^s_{ab}(x,y):=\frac{i}{2}\left(\langle\psi^+,E_A\rangle(x)\gamma^{\ A}_{(a\ |B|}
\langle E^B,e_{b)}^{\mu}\nabla_{\mu}\psi\rangle(y)
-\langle e_{(a}^{\mu}\nabla_{|\mu}\psi^+,E_{A|}\rangle(x)\gamma^{\ A}_{b)\ B}
\langle E^B,\psi\rangle(y)\right),
\]
reduces to $T_{ab}:=e^{\mu}_ae^{\nu}_bT_{\mu\nu}$ in the limit $y\rightarrow x$.
Performing a partial integration, $\int\nabla_{\mu}(e_a^{\mu}\langle v,u\rangle)=0$,
we can write $T^s_{ab}$ as a bidistribution of scalar test-functions $h_1,h_2$,
\begin{eqnarray}\label{SEMT2}
T^s_{ab}(h_1,h_2)&=&\frac{i}{2}\left(-\psi^+(E_Ah_1)\gamma^{\ A}_{(a\ |B}
\psi(\nabla_{\mu|}(E^Be_{b)}^{\mu}h_2))\right.\nonumber\\
&&+\left.\psi^+(\nabla_{\mu}(e_{(a}^{\mu}E_{|A|}h_1))\gamma^{\ A}_{b)\ B}
\psi(E^Bh_2)\right).
\end{eqnarray}
Equation (\ref{SEMT2}) can be promoted to the quantised case by replacing $\psi$ and
$\psi^+$ by the components $\psi_{SM}$ and $\psi^+_{SM}$ of the corresponding locally
covariant quantum field. The expression (\ref{SEMT2}) can be viewed as a formal
expression for the same distribution with quantised field operators.

\begin{proposition}\label{SEMTresult}
For all $f\in\Test_0(DM\oplus D^*M)$ and $h\in\Test_0(M)$ we have:
\[
\int_M \left[T^s_{ab}(x,x),B_{SM}(f)\right]h(x)d\mathrm{vol}_g(x)=
\frac{1}{2}\left\{(\nabla_{(a}B_{SM})(\gamma_{b)}(SRf) h)-
B_{SM}(\gamma_{(b}\nabla_{a)}(SRf) h)\right\},
\]
where $\nabla_a:=e^{\mu}_a\nabla_{\mu}$.
\end{proposition}
\begin{proof*}
For $f=u\oplus v$ we use Proposition \ref{prop_psi} to obtain:
\begin{eqnarray}
\left\{B_{SM}(f),\psi_{SM}^+(E_Ah)\right\}&=&-i\int_M\langle v,SE_Ah\rangle I
=i\int_M\langle S_cv,E_A\rangle hI\nonumber\\
\left\{B_{SM}(f),\psi_{SM}(\nabla_{\mu}E^Be^{\mu}_bh)\right\}&=&
-i\int_M\langle \nabla_{\mu}E^Be^{\mu}_bh,Su\rangle I=i\int_M\langle E^B,
e^{\mu}_b\nabla_{\mu}Su\rangle hI\nonumber\\
\left\{B_{SM}(f),\psi^+_{SM}(\nabla_{\mu}e^{\mu}_aE_Ah)\right\}&=&
-i\int_M\langle v,S\nabla_{\mu}e^{\mu}_aE_Ah\rangle I
=-i\int_M\langle e^{\mu}_a\nabla_{\mu}S_cv,E_A\rangle h I\nonumber\\
\left\{B_{SM}(f),\psi(E^Bh)\right\}&=&-i\langle E^B,Su\rangle hI.\nonumber
\end{eqnarray}
With equation (\ref{SEMT2}), the commutation relations and
$\left[AB,C\right]=A\left\{B,C\right\}-\left\{A,C\right\}B$ this implies
\begin{eqnarray}
\left[T^s_{ab}(x,y),B_{SM}(f)\right]&=&
\frac{1}{2}\left\{\psi^+_{SM}(E_A(x))\gamma^{\ A}_{(a\ |B|}
\langle E^B,\nabla_{b)}Su\rangle(y)\right.\nonumber\\
&&+\langle S_cv,E_A\rangle(x)\gamma^{\ A}_{(a\ |B|}
(\nabla_{b)}\psi_{SM})(E^B(y))\nonumber\\
&&-(\nabla_{(a}\psi_{SM}^+)(E_{|A|}(x))\gamma^{\ A}_{b)\ B}
\langle E^B,Su\rangle(y)\nonumber\\
&&-\left.\langle \nabla_{(a}S_cv,E_{|A|}\rangle(x)
\gamma^{\ A}_{b)\ B}\psi_{SM}(E^B(y))\right\}.\nonumber
\end{eqnarray}
In this expression we are multiplying distributions with smooth functions,
so we may take the coincidence limit yielding:
\begin{eqnarray}
\left[T^s_{ab}(x,x),B_{SM}(f)\right]&=&\frac{1}{2}\left\{
\psi_{SM}^+(\gamma_{(a}\nabla_{b)}(Su)(x))
+\nabla_{(b}\psi_{SM}(S_cv\gamma_{a)}(x))\right.\nonumber\\
&&\left.
-\nabla_{(a}\psi_{SM}^+(\gamma_{b)}Su(x))
-\psi_{SM}(\nabla_{(a}(S_cv)\gamma_{b)}(x))\right\}\nonumber\\
&=&\frac{-1}{2}\left\{\nabla_{(a}B_{SM}(\gamma_{b)}SRf(x))
-B_{SM}(\gamma_{(b}\nabla_{a)}(SRf)(x))\right\},\nonumber
\end{eqnarray}
from which the result follows.
\end{proof*}

This result can be written for spinors and cospinors separately as:
\begin{eqnarray}
\int_M \left[T^s_{ab}(x,x),\psi_{SM}(v)\right]h(x)d\mathrm{vol}_g(x)&=&
\frac{1}{2}\left\{\nabla_{(a}\psi_{SM}((S_cv)\gamma_{b)}h)
-\psi_{SM}(\nabla_{(a}(S_cv)\gamma_{b)}h)\right\}\nonumber\\
\int_M \left[T^s_{ab}(x,x),\psi_{SM}^+(u)\right]h(x)d\mathrm{vol}_g(x)&=&
\frac{-1}{2}\left\{\nabla_{(a}\psi_{SM}^+(\gamma_{b)}Suh)
-\psi_{SM}^+(\gamma_{(a}\nabla_{b)}(Su)h)\right\}.\nonumber
\end{eqnarray}

\subsubsection{Relative Cauchy evolution}\label{sssec_RCE}

To compute the relative Cauchy evolution explicitly we first note that
the isomorphism $\beta_g$ can be characterised in terms of its action on the
generators $B_{M_0}(f)$ of $\mathcal{F}_{M_0}$ as follows:
\begin{proposition}\label{RCE}
For $f\in\Test_0(DN^+\oplus D^*N^+)$ we have $\beta_g B_0(f)=B_0(T_gf)$,
where
\[
T_gf=P_g\phi_+S_gP_0\phi_-S_0f.
\]
Here the subscripts on $B$, $P$ and $S$ indicate whether they are the
objects defined on $M_0$ or $M_g$ and the smooth functions
$\phi_{\pm}$ are such that $\phi_{\pm}\equiv 1$ to the past of some
Cauchy surface in $N^{\pm}$ and $\phi_{\pm}\equiv 0$ to the future of
some other Cauchy surface in $N^{\pm}$.
\end{proposition}
\begin{proof*}
Note that $\beta^-_g\circ(\beta^-_0)^{-1}B_0(\tilde{f})=B_g(\tilde{f})$
for any $\tilde{f}\in\Test_0(DN^-\oplus D^*N^-)$. Similarly, for
$f'\in\Test_0(DN^+\oplus D^*N^+)$ we have
$\beta^+_0\circ(\beta^+_g)^{-1}B_g(f')=B_0(f')$.
The functions $\phi_{\pm},1-\phi_{\pm}$ have been chosen appropriately
in order to apply equation (\ref{timeslice1}) in Proposition
\ref{prop_LCQFTprop}. We then have $B_0(\tilde{f})=B_0(f)$, where
$\tilde{f}:=-P_0\phi_-S_0f$. Notice that $\tilde{f}$ indeed has a compact
support in $N^-$. Similarly, $B_g(\tilde{f})=B_g(f')$, where
$f':=-P_g\phi_+S_g\tilde{f}$ has support in $N^+$. Hence, for $f'=T_gf$:
$\beta_g B_0(f)=\beta_g B_0(\tilde{f})=
\beta^+_0\circ(\beta^+_g)^{-1}B_g(\tilde{f})
=\beta^+_0\circ(\beta^+_g)^{-1}B_g(f')=B_0(f')$.
\end{proof*}

On each spin spacetime
$M_{\epsilon}=(\mathcal{M},g_{\epsilon},SM_0,\pi_{\epsilon})$ we can now
quantise the Dirac field and obtain relative Cauchy evolutions
$\beta_{\epsilon}:=\beta_{g_{\epsilon}}$ on $\mathcal{F}_{N^+}$ as
before.
\begin{proposition}\label{varRCE}
Writing $\delta:=\partial_{\epsilon}|_{\epsilon=0}$ we have for all
$f\in\Test_0(DN^+\oplus D^*N^+)$:
\[
\delta(\beta_{\epsilon}B_0(f))=B_0(R(\delta \dirop_{\epsilon})S_0f).
\]
\end{proposition}
\begin{proof*}
Using the fact that $B_0$ is a $C^*$-algebra-valued distribution and
Proposition \ref{RCE} we find:
\begin{eqnarray}
\delta(\beta_{\epsilon}B_0(f))&=&\delta(B_0(P_{\epsilon}\phi_+
S_{\epsilon}P_0\phi_-S_0f))
=B_0(\delta(P_{\epsilon}\phi_+S_{\epsilon})P_0\phi_-S_0f)\nonumber\\
&=&B_0(\delta(P_{\epsilon})\phi_+S_0P_0\phi_-S_0f)
+B_0(P_0\phi_+\delta(S_{\epsilon})P_0\phi_-S_0f).\nonumber
\end{eqnarray}
Now, because $P_0\phi_-S_0f\in\Test_0(DN^-\oplus D^*N^-)$ we see that
$\delta(S_{\epsilon})P_0\phi_-S_0f$ vanishes on $J^-(N^-)$ and that
$\phi_+\delta(S_{\epsilon})P_0\phi_-S_0f$ has compact support.
Because $B_0$ solves the Dirac equation we conclude that the second term
vanishes. The first term can be rewritten using equation
(\ref{timeslice1}), which yields
$S_0f=-S_0P_0(\phi_-S_0f)$ and hence:
\[
\delta(\beta_{\epsilon}B_0(f))=-B_0(\delta(P_{\epsilon})\phi_+S_0f)
=-B_0(\delta(P_{\epsilon})S_0f).
\]
For the last equality we used the fact that $\delta(P_{\epsilon})$ is
supported in $K$, where $\phi_+\equiv 1$. Recall that
$P=(-i\dirop+m)\oplus(i\dirop+m)$ to get the final result.
\end{proof*}

To compute the variation of the Dirac operator we may work in a local
frame on $O$, where it is supported. Because the Dirac adjoint map is
independent of $\epsilon$ we only need to compute this variation either
for spinors or for cospinors:
\begin{lemma}\label{adjointvar}
For $v\in\Test_0(D^*M)$ we have
$\delta(\dirop)v=(\delta(\dirop)v^+)^+$.
\end{lemma}
\begin{proof*}
Because the adjoint operation is continuous we have:
\[
\delta(\dirop) v=\partial_{\epsilon}\dirop_{\epsilon}v|_{\epsilon=0}
=\partial_{\epsilon}(\dirop_{\epsilon}v^+)^+|_{\epsilon=0}
=(\partial_{\epsilon}\dirop_{\epsilon}v^+|_{\epsilon=0})^+
=(\delta(\dirop)v^+)^+.
\]
\end{proof*}

It is interesting to note that only the variation of the Dirac operator
is of importance for the variation of the relative Cauchy evolution, just
like for the stress-energy-momentum tensor (cf.\ \cite{Forger+}). It will
also turn out that the variation only depends on the variation of the
metric and not on the other freedom in the variation of the orthonormal
frame, even though we are now acting on it with the $C^*$-algebra-valued
field (cf.\ Remark \ref{rem_variation}). This will follow from the proof
of the following theorem, for which we refer to appendix \ref{sec_proof}.
\begin{theorem}\label{thm_RCE}
For a double test-spinor $f\in\Test_0(DM_0\oplus D^*M_0)$ and $x\in K$:
\begin{eqnarray}
\frac{\delta}{\delta g^{\alpha\beta}(x)}\left(\beta_gB_0(f)\right)&=&
-B_0\left(\frac{\delta}{\delta g^{\alpha\beta}(x)} P_gS_0f\right)=
\frac{-i}{2}e^a_{\alpha}e^b_{\beta}\left[T^s_{ab}(x,x),B_0(f)\right].
\end{eqnarray}
\end{theorem}
This result compares well with the scalar field case, Theorem 4.3 in
\cite{Brunetti+}.\footnote{The sign explained in the footnote on page
\pageref{ft_RCEsign} cancels the sign due to the variation
w.r.t.\ $g^{\alpha\beta}$ instead of $g_{\alpha\beta}$.}
As particular cases we obtain for $\psi$ and $\psi^+$:
\begin{eqnarray}
\frac{\delta}{\delta g^{\alpha\beta}(x)}(\beta_g\psi(v))&=&
\frac{-i}{2}e^a_{\alpha}e^b_{\beta}\left[T^s_{ab}(x,x),\psi(v)\right],
\nonumber\\
\frac{\delta}{\delta g^{\alpha\beta}(x)}(\beta_g\psi^+(u))&=&
\frac{-i}{2}e^a_{\alpha}e^b_{\beta}\left[T^s_{ab}(x,x),\psi^+(u)\right].
\nonumber
\end{eqnarray}
It follows that the same result also holds for products and sums of
smeared field operators.

\section{Conclusions}

A rigorous formulation of quantum field theories in curved spacetime, going
beyond the well-known scalar field, is a prerequisite for constructing
more realistic cosmological models as well as for improving our understanding
of quantum field theory in Minkowski spacetime. The main purpose of this
paper was to present the free Dirac field in a four dimensional globally
hyperbolic spacetime as a locally covariant quantum field theory in the sense
of \cite{Brunetti+} and to compute the relative Cauchy evolution of this field,
obtaining commutators with the stress-energy-momentum tensor in analogy with
the free real scalar field. We achieved this in a representation independent
way and in a functorial, and therefore manifestly covariant, framework.

We established some basic properties of the locally covariant free Dirac
field and remarked on the quantisation of Majorana spinors. We also provided
a detailed discussion of Hadamard states, closing any gaps in the existing
proofs of the equivalence of the definitions in terms of the series expansion
of their two-point distribution and a microlocal condition, respectively.

Furthermore we argued that the observable part of the theory is uniqueley
determined by the relations between adjoints, charge conjugation and the Dirac
operator, although the geometric constructions themselves may not be unique due
to the cohomological properties of the category of spin spacetime. On a
mathematical level we have consistently replaced a single spin spacetime
$SM$ by the category $\cat{SSpac}$ of such spacetimes, and the differential
geometry on $SM$ by the corresponding functorial descriptions. On a physical
level, however, we should not conclude from this that $\cat{SSpac}$ is now
the physical arena in which our system lives, instead of a collection of
systems. (See Ch.1 of \cite{KS2} for more detailed philosophical remarks on
the interpretation of the locally covariant approach.)

${}$\\[15pt]
{\bf Acknowledgements}\\[6pt]
I would like to thank Chris Fewster for suggesting to use the cohomological
language in subsection \ref{ssec_uniqueness} and for bringing the problem
of computing the relative Cauchy evolution for the Dirac field to my
attention. I would also like to thank Romeo Brunetti for correcting some of
my misconceptions in the early stages of this computation. Much of this
work was performed as part of my PhD-thesis at the University of York and
I would also like to thank the University of Trento for their kind
hospitality during my visit in October 2007. Furthermore this research was
supported by the German Research Foundation (Deutsche Forschungsgemeinschaft
(DFG)) through the Institutional Strategy of the University of G\"ottingen
and the Graduiertenkolleg 1493 ''Mathematische Strukturen in der modernen
Quantenphysik''.

\appendix
\section{Results on microlocal analysis}\label{sec_ma}

In this appendix we will list some results concerning the microlocal analysis
of distributions. For a detailed treatment of scalar distributions we refer to
\cite{Hoermander}, whereas Hilbert and Banach-space-valued distributions are
treated in \cite{Strohmaier+,KS2}. More details concerning distributional
sections of vector bundles can be found in e.g.\ \cite{Baer+,Sahlmann+,Dencker,KS2}.

Before we discuss distributional sections of vector bundles we first consider
the scaling limit of a distribution in an open set of $\mathbb{R}^n$:
\begin{definition}
Let $O$ be a convex open region $O\subset\mathbb{R}^n$ containing $0$. For all
$\lambda>0$ we define the \emph{scaling map}
$\map{\delta_{\lambda}}{O}{O}$ by $\delta_{\lambda}(x):=\lambda x$.

Let $u$ be a distribution on a convex open region $O\subset\mathbb{R}^n$
containing $0$. The \emph{scaling degree} $d$ of $u$ at $0$ is defined as
$d:=\inf\left\{\beta\in\left[-\infty,\infty\right)|\ 
\lim_{\lambda\rightarrow 0}\lambda^{\beta}\delta_{\lambda}^*u=0\right\}$,
where $(\delta_{\lambda}^*u)(f):=\lambda^{-n}u(f\circ\delta_{\lambda}^{-1})$.

If $u^0:=\lim_{\lambda\rightarrow 0}\lambda^d\delta_{\lambda}^*u$ exists we
call it the \emph{scaling limit} of $u$ at $0$.
\end{definition}
Note that the scaling limit may fail to exist (e.g.\ $u(x)=\log|x|$) or it
may vanish (e.g.\ if $0\not\in\mathrm{supp}(u)$). On a manifold, we will only
consider scaling limits in a certain choice of local coordinates. How this
limit depends on this choice of coordinates will not be relevant for us.

We now prove the following result:\footnote{A similar result was also claimed
in \cite{Sahlmann+}, but we find the proof unconvincing. In particular, when
localising the scaling limit with a test-function $\chi_0$ and estimating
\[
\widehat{\chi_0u^0}(\xi)=\lim_{\lambda\rightarrow 0}\lambda^{d-n}
u\left(\chi_0\left(\frac{.}{\lambda}\right)e^{-i\frac{\xi}{\lambda}\cdot .}\right)
\]
the test-function $\chi_0(\frac{.}{\lambda})$ becomes singular in the limit
$\lambda\rightarrow 0$. The quoted reference pays insufficient attention to
this issue.}\label{ft_scaling}
\begin{proposition}\label{prop_scale}
Let $u$ be a distribution on a convex open region $O\subset\mathbb{R}^n$
containing $0$ with scaling limit $u^0$ at $0$. Then
\[
\left\{0\right\}\times\pi_2(WF(u^0))\subset WF(u),
\]
where $\pi_2$ denotes the projection on the second coordinate.
\end{proposition}
\begin{proof*}
Suppose that $(0,\xi_0)\not\in WF(u)$ with $\xi_0\not=0$. We will prove that
$(x,\xi_0)\not\in WF(u^0)$ for all $x$. By assumption we can choose
$\chi\in\Test_0(O)$ and an open conic neighbourhood $\Gamma\subset\mathbb{R}^n$
of $\xi_0$ such that $\chi\equiv 1$ on a neighbourhood of $0$ and
$\mathrm{supp}(\chi)\times\Gamma\cap WF(u)=\emptyset$. We set $v:=\chi u$
and $v^{\lambda}:=\lambda^d\delta_{\lambda}^*v$, where $d$ is the scaling
degree of $u$ at $0$. Notice that $WF(v)\cap T^*_0O=WF(u)\cap T^*_0O$ and 
$u^0:=\lim_{\lambda\rightarrow 0}v^{\lambda}$, so without
loss of generality we may prove the result with $v$ replacing $u$ and we
can view the $v^{\lambda}$ as compactly supported distributions on all of
$\mathbb{R}^n$.

Notice that for $\lambda>0$ we have $\delta_{\lambda}^*u^0=\lambda^{-d}u^0$,
i.e.\ $u^0$ is a homogeneous distribution and therefore it is tempered
(\cite{Hoermander} Theorem 7.1.18). We now prove that $v^{\lambda}$ converges
to $u^0$ in the sense of tempered distributions on $\mathbb{R}^n$. For this
we first write
$v=\sum_{|\alpha|\le r}(-1)^{|\alpha|}\partial^{\alpha}v_{\alpha}$, where
$r$ is the order of $v$ and the $v_{\alpha}$ are compactly supported
distributions of order $0$ (see \cite{Hoermander} Section 2.1). Note that
$\sum_{|\alpha|<d-n}(-1)^{|\alpha|}\partial^{\alpha}v_{\alpha}$
converges to $0$ in $\mathcal{S}$, because for every $|\alpha|<d-n$ and
$\phi\in\mathcal{S}(\mathbb{R}^n)$ we have
\[
|((-1)^{|\alpha|}\partial^{\alpha}v_{\alpha})^{\lambda}(\phi)|=
\lambda^{d-n}|v_{\alpha}(\partial^{\alpha}(\phi\circ\delta_{\lambda}^{-1}))|
\le\lambda^{d-n-|\alpha|}C\sup|\partial^{\alpha}\phi|
\]
which converges to $0$ as $\lambda\rightarrow 0$. We then set
$w:=\sum_{d-n\le |\alpha|\le r}(-1)^{|\alpha|}\partial^{\alpha}v_{\alpha}$, so
that $\lim_{\lambda\rightarrow 0}w^{\lambda}=u^0$ as distributions. By the
Uniform Boundedness Principle this implies
\begin{equation}\label{eqn_uniformest}
|w^{\lambda}(\phi)|\le C\sum_{|\alpha|\le r}\sup|\partial^{\alpha}\phi|
,\quad \mathrm{supp}(\phi)\subset B_1,
\end{equation}
for some $C,r>0$, where $B_1$ is the (Euclidean) unit ball and 
$0<\lambda\le 1$. In fact, for $\lambda\ge 1$ we also have
\[
|w^{\lambda}(\phi)|=\lambda^{d-n}|w(\phi\circ\delta_{\lambda}^{-1})|
\le C\sum_{d-n\le|\alpha|\le r}\lambda^{d-n-|\alpha|}\sup 
|\partial^{\alpha}\phi|\le C\sum_{d-n\le|\alpha|\le r}\sup |\partial^{\alpha}\phi|,
\]
so the estimate (\ref{eqn_uniformest}) holds for all $\lambda>0$.

Now let $\phi\in\mathcal{S}(\mathbb{R}^n)$ be a function of rapid decrease and
choose a partition of unity on $\mathbb{R}^n$ as follows. We let
$\chi_0\in\Test_0(\mathbb{R}^n)$ be positive such that $\chi\equiv 1$ on $B_1$
and $\chi(x)=0$ when $\|x\|\ge 2$. We then set
$\chi_m(x):=\chi_0(2^{-m}x)-\chi_0(2^{1-m}x)$ and note that:
\[
\mathrm{supp}(\chi_{m\ge 1})\subset\left\{x|\ 2^{m-1}\le\|x\|\le 2^{m+1}\right\}
,\quad\sum_{m=0}^{\infty}\chi_m=1,
\]
where the sum is finite near every point. We define $\phi_m:=\chi_m\phi$ and
$\mu_m:=2^{-m-1}$ and rescale $\phi_m$ in order to apply the estimate
(\ref{eqn_uniformest}):
\begin{eqnarray}\label{eqn_est2}
|w^{\lambda}(\phi_m)|&=&\mu_m^{d-n}\left|w^{\lambda/\mu_m}\left(\phi_m\left(
\frac{.}{\mu_m}\right)\right)\right|
\le C\sum_{|\alpha|\le r}\mu_m^{d-n-|\alpha|}\sup
\left|(\partial^{\alpha}\phi_m)\left(\frac{.}{\mu_m}\right)\right|\nonumber\\
&\le&
C_1\sum_{|\alpha|\le r}\sum_{|\beta|\le r+n-d}\sup_{\mathbb{R}^n}
|x^{\beta}\partial^{\alpha}\phi_m|,\quad m\ge 0,
\end{eqnarray}
where the last line uses $\mu_m^{d-n-|\alpha|}\le (4\|x\|)^{|\alpha|+n-d}$ for
$m\ge 1$, which follows from $d-n\le|\alpha|$ and the support properties of
$\chi_m$. (For $m=0$ we simply estimate $\mu_0^{d-n-|\alpha|}$ by a constant to
arrive at the last line of (\ref{eqn_est2}.)
We now note that $\max_{\alpha}\sup_x|\partial^{\alpha}\chi_m|\le c$ for
some $c$ independent of $m$, as the derivatives only bring out extra factors
of $2^{-m}\le 1$. Moreover, for $m\ge 0$ we notice that
$\chi_{m+1}+\chi_m+\chi_{m-1}\equiv 1$ on $\mathrm{supp}(\chi_m)$, where
we define $\chi_{-1}:=0$. Therefore (\ref{eqn_est2}) leads to
\[
|w^{\lambda}(\phi_m)|\le C_2\sum_{|\alpha|\le r}\sum_{|\beta|\le r+n-d}
\sup_{\mathbb{R}^n}|x^{\beta}\partial^{\alpha}\phi|(\chi_{m+1}+\chi_m+\chi_{m-1})
\]
and summing over $m\ge 0$ then gives:
\[
|w^{\lambda}(\phi)|\le 3C_2\sum_{|\alpha|\le r}\sum_{|\beta|\le r+n-d}
\sup_{\mathbb{R}^n}|x^{\beta}\partial^{\alpha}\phi|.
\]
This shows that $w^{\lambda}(\phi)$ can be estimated by a seminorm on
$\mathcal{S}(\mathbb{R}^n)$ uniformly in $\lambda$. It then follows that
$w^{\lambda}\rightarrow u^0$ and hence $v^{\lambda}\rightarrow u^0$ as tempered
distributions. Indeed, for any $\phi\in\mathcal{S}(\R^n)$ and $\epsilon>0$
we can choose $\phi'\in\Test_0(\R^n)$ and $\lambda_0>0$ such that
$|w^{\lambda}(\phi-\phi')|<\frac{\epsilon}{2}$ for all $\lambda>0$ and
$|w^{\lambda}(\phi')|<\frac{\epsilon}{2}$ for all $\lambda<\lambda_0$.

Fourier transformation is a continuous operation on tempered distributions, so
we can compute:
\begin{eqnarray}
|\widehat{u^0}(\xi)|&=&\lim_{\lambda\rightarrow 0}\lambda^{d-n}
\left|\hat{v}\left(\frac{\xi}{\lambda}\right)\right|\le
C_N\lim_{\lambda\rightarrow 0}\lambda^{d-n} \left\|\frac{\xi}{\lambda}\right\|^{-N}
=C_N\|\xi\|^{-N}\lim_{\lambda\rightarrow 0}\lambda^{N+d-n}\nonumber
\end{eqnarray}
for all $\xi$ in $\Gamma$, all $N\in\mathbb{N}$ and suitable $C_N> 0$. For
$N> n-d$ the limit yields $\widehat{u^0}(\xi)=0$ near $\xi_0$. We then apply
Theorem 8.1.8 in \cite{Hoermander}, which says that for a homogeneous
distribution we have for all $x\not=0$ that $(x,\xi_0)\in WF(u^0)$ if and only if
$(\xi_0,-x)\in WF(\widehat{u^0})$ and also $(0,\xi_0)\in WF(u^0)$ if and only if $\xi_0\in\mathrm{supp}(\widehat{u^0})$.
\end{proof*}

For a distribution $u$ with values in a Banach space $\mathcal{B}$ one can
define the wave front set by using estimates of the norm
$\|u(\chi e^{i\xi\cdot})\|$, which replace the corresponding estimates of the
absolute value $|u(\chi e^{i\xi\cdot})|$ for scalar distributions \cite{Strohmaier+}.
Alternatively, one can use the following equivalent characterisation 
(\cite{KS2}, Theorem A.1.4):
\begin{equation}\label{eqn_WFBvalued}
WF(u)=\overline{\cup_{l\in\mathcal{B}'}WF(l\circ u)}\setminus\mathcal{Z}.
\end{equation}
A similar idea works for a distributional section $u$ of a vector bundle
$\mathcal{V}=O\times\mathbb{R}^m$ over a contractible region $O$ of $\mathbb{R}^n$.
Indeed, using a basis $e_i$ for $\mathbb{R}^m$ with dual basis $e^i$ we can
identify $u$ with a distribution $\tilde{u}$ on $O$ with values in
$\mathcal{B}\otimes(\mathbb{R}^m)'$, where the correspondence is given by
\[
\tilde{u}(h):=\sum_{i=1}^mu(he_i)\otimes e^i,\quad
u\left(\sum_{i=1}^mf^ie_i\right)=\sum_{i=1}^m\langle\tilde{u}(f^i),e_i\rangle,
\]
where $\langle,\rangle$ denotes the canonical pairing of $\mathbb{R}^m$ with the
second factor of $\mathcal{B}\otimes(\mathbb{R}^m)'$. We set by definition
$WF(u):=WF(\tilde{u})$.

Equation (\ref{eqn_WFBvalued}) allows a straightforward generalisation of
many results for scalar distributions on open sets of $\mathbb{R}^n$ to
Banach-space-valued distributional sections of a vector bundle over regions over
$\mathbb{R}^n$. Moreover, by showing how these results transform under changes
of coordinates they can be formulated for vector bundles on a manifold. We list
a number of these results in the following Theorem (cf.\ \cite{Hoermander,KS2}):
\begin{theorem}\label{thm_WFBvalued}
If $u,v$ are distributional sections of a complex vector bundle $\mathcal{V}$ over
the spacetime $M$ with values in the Banach space $\mathcal{B}$, then:
\begin{enumerate}
\item $\mathrm{sing\ supp}(u)$ is the projection of $WF(u)$ on the first variable,
\item $u\in\Test(\mathcal{V},\mathcal{B})$ if and only if $WF(u)=\emptyset$,
\item $WF(u+v)\subset WF(u)+WF(v)$,
\item if $P$ is a linear partial differential operator on $\mathcal{V}$ with
smooth coefficients and (matrix-valued) principal symbol\footnote{See \cite{Baer+}
for the definition of the principal symbol.} $p(x;\xi)$, then
$WF(Pu)\subset WF(u)\subset WF(Pu)\cup\Omega_P$, where
$\Omega_P:= \left\{(x;\xi)\in T^*M|\ \xi\not=0,\ \det p(x;\xi) = 0\right\}$,
\item if $x\in M$, $\map{\phi}{U}{\mathbb{R}^n}$ is a local trivialisation on a
convex neighbourhood $U$ with $\phi(x)=0$ and $(\phi^{-1})^*u$ has a scaling limit
$u^0$ at $0$, then
$\phi^*(\left\{0\right\}\times\pi_2(WF(u^0)))\subset WF(u)\cap T^*_xM$.
\end{enumerate}
\end{theorem}
In the last item, the scaling limit depends not just on the choice of coordinates,
but also on the choice of a frame $e_i$ of $\mathcal{V}$ over $U$ and we let the
scaling maps $\delta_{\lambda}$ act on sections of $\mathcal{V}$ componentwise:
$(\sum_if^ie_i)\circ\delta_{\lambda}^{-1}=
\sum_i(f^i\circ\delta_{\lambda}^{-1})e_i$.

In the particular case where $\mathcal{B}$ is a Hilbert space we also have
(see \cite{Strohmaier+,KS2}):
\begin{theorem}\label{WFHvalued}
Let $\mathcal{H}$ be a Hilbert space and $\mathcal{V}_i$, $i=1,2$, two finite
dimensional (complex) vector bundles over smooth $n_i$ dimensional spacetimes
$M_i$ with complex conjugations $J_i$, i.e.\ the $J_i$ are antilinear, base-point
preserving bundle isomorphisms $\map{J_i}{\mathcal{V}_i}{\mathcal{V}_i}$ such that
$J^2_i=-id$. Let $u_i$, $i=1,2$, be two $\mathcal{H}$-valued distributional
sections of $\mathcal{V}_i$ and let $w_{ij}$ be the distributional sections of the
vector bundle $\mathcal{X}_i\boxtimes\mathcal{X}_j$ over $M_i\times M_j$ determined
by $w_{ij}(f_1\boxtimes f_2):=\langle u_i(J_if_1), u_j(f_2)\rangle$. Then
\[
(x,\xi)\in WF(u_1)\quad\Leftrightarrow (x,-\xi; x,\xi)\in WF(w_{11})
\]
and
\[
WF(w_{ij})\subset -(WF(u_i)\cup\mathcal{Z})\times (WF(u_j)\cup\mathcal{Z}),
\]
where $\mathcal{Z}$ denotes the zero-section.
\end{theorem}

Finally we establish some results on the wave front sets of advanced and retarded
fundamental solutions $E^{\pm}$ (for their existence and uniqueness we refer to
\cite{Baer+}) and $S^{\pm},S_c^{\pm}$. These results are analogous to Theorem
6.5.3 of \cite{Duistermaat+}, but now for operators in a vector bundle. Note that
for distributional sections of vector bundles there is a Propagation of
Singularities Theorem, which follows from the propagation of the polarisation set
\cite{Dencker}.
\begin{theorem}\label{WFE}
Let $E^{\pm}$ be the advanced ($-$) and retarded ($+$) fundamental
solutions for a normally hyperbolic operator $P$ acting on the sections
of a vector bundle $DM$ over a globally hyperbolic spacetime
$M=(\mathcal{M},g)$ of dimension $n\ge 2$. Then
\begin{eqnarray}\label{eqn_WFE}
WF(E^{\pm})&=&\left\{(x,\xi;y,\eta)\in T^*M^{\times 2}\setminus\mathcal{Z}|
x\in J^{\pm}(y),\ x\not=y,\ (x,-\xi)\sim(y,\eta)\right\}\nonumber\\
&&\cup\left\{(x,-\xi;x,\xi)\in T^*M^{\times 2}\setminus\mathcal{Z}|
(x,\xi)\in T^*M\setminus\right\}\nonumber\\
&=:&A^{\pm}\cup B
\end{eqnarray}
where $\mathcal{Z}$ is the zero-section and $(x,\xi)\sim(y,\eta)$ if and
only if there is a light-like geodesic $\gamma$ from $x$ to $y$ to which
$\xi$ and $\eta$ are cotangent such that they are each others parallel
transport along $\gamma$.
\end{theorem}
\begin{proof*}
The first part of this proof follows closely the proof of \cite{Radzikowski}.

We start by reducing the problem to a local one as follows. The principal symbol
of $P$ is $p(x,\xi)=g_{\mu\nu}(x)\xi^{\mu}\xi^{\nu}I$, where $I$ is the identity
operator on $DM$, so by the Propagation of Singularities Theorem, the
singularities of $E^{\pm}$ propagate along light-like geodesics by parallel
transport. By definition the points in set $A^{\pm}$ are invariant under the
same parallel transport. Now consider a point $p:=(x,\xi;y,\eta)$ with $x\not=y$.
If $\xi=\eta=0$ then $P$ is not contained in any set on either side of the
equality, so we may assume $\xi\not=0$ (the case $\eta\not=0$ is analogous).
Let $S$ be a spacelike Cauchy surface through $y$ and propagate
$(x,\xi)$ along the light-like geodesic $\gamma$ towards $S$. If $\gamma$
ends at $S$ in $x'\not=y$ then $P$ is not contained in $A^{\pm}$ or $B$,
nor is it contained in $WF(E^{\pm})$, because $E(x',y)=0$ when $x'$ and $y$
are spacelike, so it cannot have any singularities there. If $\gamma$ ends at
$y$, on the other hand, we can find a point $p':=(x',\xi';y,\eta)$, where $x'$
on $\gamma$ is in any given causally convex neighbourhood of $y$ and $\xi'$ is
the parallel transport of $\xi$ along $\gamma$ to $x'$. Then
$p'\in WF(E^{\pm})$ if and only if $p\in WF(E^{\pm})$ and $p'\in A^{\pm}$ if
and only if $p\in A^{\pm}$. Hence, it suffices to prove the claim locally.

On a sufficiently small causally convex domain $O\subset\mathcal{M}$ we
can find for every $k\in\mathbb{N}$ a $C^k$-section $W^k$ of
$DM\boxtimes D^*M$ on $O^{\times 2}$ such that (\cite{Baer+}
Proposition 2.5.1):
\begin{equation}\label{eqn_Hadseries}
E^{\pm}(x,y)=\sum_{j=0}^{k+1}V_j(x,y)f^*(1\otimes R^{\pm}(2+2j,.))(x,y)+
W^k(x,y).
\end{equation}
Here the Hadamard coefficients $V_j$ are uniquely defined smooth sections
of $DM\boxtimes D^*M$ on $O^{\times 2}$, $R^{\pm}(\alpha,y)$ are the
retarded ($+$) and advanced ($-$) Riesz distributions (or rather distribution
densities) on Minkowski spacetime and they are pulled back by the smooth
diffeomorphism $\map{f}{O^{\times 2}}{TO}$ defined by
$(x,y)\mapsto (x,\exp_x^{-1}(y))$. This means we use Riemannian normal
coordinates for $y$ centered on $x$, which is well-defined because $O$ is
causally convex. The Riesz distributions have many useful properties, of
which we will only use for all $j\ge 0$:
\begin{eqnarray}\label{eqn_propR}
WF(R^{\pm}(2j+2,.))&=&\left\{(x,\xi)\in T^*M_0\setminus\mathcal{Z}|\
x=0\mathrm{\ or\ } x^2=0, x\in J^{\pm}(0), \xi\parallel x\right\}\nonumber\\
R^{\pm}(2+2j,\lambda x)&=&\lambda^{2+2j-n}R^{\pm}(2+2j,x),\ \lambda>0
\end{eqnarray}
(These can be proved using \cite{Baer+} Proposition 1.2.4 items 4, and 5, 
$\Box^{j+1}R^{\pm}(2+2j,.)=\delta$ and the wave front sets of the distinguished
parametrices as determined in \cite{Duistermaat+}.) Hence, for all $j\in\mathbb{N}$:
\begin{eqnarray}\label{eqn_WFfR}
WF(f^*(1\otimes R^{\pm}(2+2j,.)))&=&f^*(WF(1\otimes R^{\pm}(2+2j,.)))
=f^*(\mathcal{Z}|_O\times WF(R^{\pm}(2+2j,.)))\nonumber\\
&=&\left\{(x,\xi;y,\eta)|\ (\xi,\eta)=df^T(0,\eta')\ \mathrm{for\ some\ }
\right.\nonumber\\
&&\left.(\exp_x^{-1}(y),\eta')\in WF(R^{\pm}(2+2j,.))\right\},\nonumber\\
&=&(A^{\pm}\cup B)\cap T^*O^{\times 2},
\end{eqnarray}
where $df^T$ is the transpose of the derivative $df$ at $(x,y)$.
The last equality uses the wave front set of the Riesz distributions in
equation (\ref{eqn_propR}) and the properties of Riemannian normal coordinates
(cf.\ \cite{Radzikowski}). It follows that
$WF(E^{\pm}|_{O^{\times 2}})\subset (A^{\pm}\cup B)\cap T^*O^{\times 2}$,
because for each order of differentiation $N$ we can choose a sufficiently high
order $k$ in equation (\ref{eqn_Hadseries}) to make the required estimate in
the definition of the wave front set.

We can prove the opposite inclusion, if we
can show that the wave front set of the finite sum in (\ref{eqn_Hadseries})
also contains $(A^{\pm}\cup B)\cap T^*O^{\times 2}$, which we will do using
scaling limits (cf.\ \cite{Sahlmann+}). First we may employ the Riemannian
normal coordinates $\map{f}{O^{\times 2}}{TO}$ as above. Next we may assume that
$O$ is also a contractible coordinate neighbourhood, so we can consider local
coordinates $\map{\phi}{O}{\mathbb{R}^n}$ on $O$ and the associated coordinate
map $d\phi$ on $TO$. Moreover, we can choose $\phi$ in such a way that
$\phi(x_0)=0$ for an arbitrarily given $x_0\in O$. The composition
$d\phi\circ f$ then defines coordinates on $O^{\times 2}$ such that
$(x_0,x_0)\mapsto 0\in\mathbb{R}^{2n}$. Using a frame $E_A$ for $DM|_O$ and the
dual frame $E^B$ we can express the terms in the sum of equation
(\ref{eqn_Hadseries}) in the local coordinates $d\phi\circ f$ as
$V_{jB}^A(x,y)R^{\pm}(2+2j,y)$. From equation (\ref{eqn_propR}) we then find
the scaling behaviour
\[
\delta_{\lambda}^*(V_{jB}^A(x,y) R^{\pm}(2+2j,y))=
\lambda^{2+2j-n}(V_{jB}^A(\lambda x,\lambda y) R^{\pm}(2+2j,y))
\]
for all $\lambda>0$. In the scaling limit only the lowest order term survives:
\begin{eqnarray}
\lim_{\lambda\rightarrow 0}\lambda^{n-2} (\delta_{\lambda}\circ f^{-1}\circ
d\phi^{-1})^* E(x,y)&=&V_{0B}^A(0,0)R(2,y)E^B(x)E_A(y)
=R(2,y)E^A(x)E_A(y),\nonumber
\end{eqnarray}
where we wrote $R(2,y):=R^-(2,y)-R^+(2,y)$ and we used the explicit expression
$V_{0\ B}^A(x,x)=\delta^A_B$ (\cite{Baer+} Lemmas 2.2.2 and 1.3.17).

Now, the last item of Theorem \ref{thm_WFBvalued} (which follows from Proposition
\ref{prop_scale}) implies that $WF(E)\supset
(d\phi\circ f)^*(\left\{(0,0)\right\}\times\pi_2(WF(1\otimes R(2,.))))$, because
$E^A(x)E_A(y)$ is smooth and not identically vanishing. From equation
(\ref{eqn_propR}) and the support properties of $R^{\pm}(2,.)$ we easily compute
$\pi_2(WF(1\otimes R(2,.)))=\left\{(0,\xi)|\ \xi^2=0\right\}$. Pulling this back
to $O^{\times 2}$ and using the properties of Riemannian normal coordinates yields
\[
WF(E)\supset\left\{(x_0,-\xi;x_0,\xi)|\ \xi^2=0\right\}.
\]
Because $E$ is a bi-solution to the wave equation we can apply the Propagation
of Singularities Theorem to find that $WF(E)\supset A^+\cup A^-$ on $O^{\times 2}$
and from the support properties of $E^+$ and $E^-$ we then conclude that
$WF(E^{\pm})\supset A^{\pm}$. Finally, $WF(E^{\pm})\supset WF(PE^{\pm})=WF(\delta)=B$.
This completes the proof.
\end{proof*}

\begin{corollary}\label{cor_WFE}
In the notation of Theorem \ref{WFE}, $WF(E)=\overline{A^+\cup A^-}\setminus\mathcal{Z}$.
\end{corollary}
\begin{proof*}
By Theorem \ref{WFE} and the support properties of $E^{\pm}$ we have
$WF(E)=A^+\cup A^-$ away from the diagonal. The inclusion $\supset$ then follows
from the closedness of the wave front set. For the opposite inclusion we consider
a point on the diagonal and use the Propagation of Singularities Theorem to find an
approximating sequence of points off the diagonal.
\end{proof*}

\begin{proposition}\label{WFS}
For the fundamental solutions of the Dirac equation we have, in the notation of
Theorem \ref{WFE}: $WF(S^{\pm})=WF(S_c^{\pm})=A^{\pm}\cup B$ and
$WF(S)=WF(S_c)=\overline{A^+\cup A^-}\setminus\mathcal{Z}$.
\end{proposition}
In other words, $WF(S^{\pm})=WF(S_c^{\pm})=WF(E^{\pm})$ and 
$WF(S)=WF(S_c)=WF(E)$.
\begin{proof*}
Because $S^{\pm}=(i\dirop+m)E^{\pm}$ and $S_c^{\pm}=(-i\dirop+m)E^{\pm}$ (see
\cite{Dimock}) we immediately find $WF(S^{\pm})\subset WF(E^{\pm})$ and
$WF(S_c^{\pm})\subset WF(E^{\pm})$. Similarly $WF(S)\subset WF(E)$ and
$WF(S_c)\subset WF(E)$. Now suppose that
$WF(S)=WF(S_c)=WF(E)=\overline{A^+\cup A^-}$, which we will prove below.
By the support properties of the fundamental solutions we then find that away
from the diagonal $WF(S^{\pm})=WF(S_c^{\pm})=A^{\pm}$, whereas on the
diagonal $WF(E^{\pm})=B\supset WF(S^{\pm})\supset WF(PS^{\pm})=WF(\delta)=B$
and similarly for cospinors.

To complete the proof we need to show that $WF(S)\supset WF(E)$ and
$WF(S_c)\supset WF(E)$, for which we adapt (and correct) an idea of
\cite{Hollands2}. We prove the case of $S$, because the other case follows
by taking adjoints (cf.\ Theorem \ref{thm_fundsln}). Further note that it is
sufficient to prove the claim on the diagonal, because the Propagation of
Singularities Theorem applies both to $E$ and to $S$. Now suppose that
$(x,-\xi;x,\xi)\in WF(E)\setminus WF(S)$. We will derive a contradiction as
follows. For every time-like, future pointing normalised vector $n_0\in T_xM$
we can find a smooth spacelike Cauchy surface $C$ through $x$ such that $n_0$
is normal to $C$. We let $n$ denote the future pointing normal vector field
on $C$ and $\map{\iota}{C}{M}$ the canonical injection. By \cite{Dimock}
Proposition 2.4c) we can restrict $S$ to $C^{\times 2}$ to find
$S|_{C^{\times 2}}=-i\delta\fsl{n}$ and in particular
$(x,-d\iota^T_x(\xi);x,d\iota^T_x(\xi))\in WF(S|_{C^{\times 2}})$. By (a
component version of) \cite{Hoermander} Theorem 8.2.4, on the other hand:
\[
WF(S|_{C^{\times 2}})\subset (\iota\times\iota)^*(WF(S))
=\left\{(x,d\iota^T_x(\xi);y,d\iota^T_y(\xi'))|\ (x,\xi;y,\xi')\in WF(S)\right\}.
\]
Therefore, there must be a point $(x,-\eta;x,\eta)\in WF(S)$ such that
$(x,-d\iota^T_x(\eta);x,d\iota^T_x(\eta))=(x,-d\iota^T_x(\xi);x,d\iota^T_x(\xi))$
Notice, however, that the transpose of $d\iota$ is nothing else than restricting
the dual vector $\xi$ to the tangent space of $C$. Because $WF(S)\subset WF(E)$
there are only two possibilities: $\eta=\xi$ or $\eta=\xi-2(\xi_an_0^a)n_0$. The
first contradicts our assumption, so we have $\eta=\xi-2(\xi_an_0^a)n_0$. Now
$(x,-\eta;x,\eta)\in WF(S)$ must hold for every normalised, time-like, future
pointing vector $n_0\in T_xM$. Choosing a sequence of vectors $n_0$ such that
$\eta\rightarrow\xi$ and using the closedness of the wave front set we find
again $(x,-\xi;x,\xi)\in WF(S)$. Hence, $WF(E)=WF(S)$.
\end{proof*}

\section{Proof of Theorem \ref{thm_RCE}}\label{sec_proof}

The computations involved in the proof of Theorem \ref{thm_RCE} are somewhat
similar to the computation of the stress-energy-momentum tensor. We will work
in components and in local coordinates on $O$, using Greek indices to indicate
the coordinate frame and coordinate derivatives. To ease the notation we will
drop the subscript $\epsilon$ on the local frame $e_a^{\mu}$.

As $\gamma^a$ is independent of $\epsilon$ we may use equations
(\ref{spinconnection}) to vary
\begin{equation}\label{var0}
\dirop v=\left(\partial_av-\frac{1}{4}\Gamma^c_{\ ab}v\gamma_c\gamma^b
\right)\gamma^a
=e_a^{\alpha}\left(\partial_{\alpha}v
+\frac{1}{4}e_b^{\beta}\left\{\partial_{\alpha}e^c_{\beta}
-e^c_{\gamma}\Gamma^{\gamma}_{\ \alpha\beta}\right\}
v\gamma_c\gamma^b\right)\gamma^a,
\end{equation}
which yields:
\begin{eqnarray}\label{var1}
\delta\dirop v&=&\delta e_a^{\alpha}e^d_{\alpha}\nabla_dv\gamma^a
-\frac{1}{4}\delta e_b^{\beta}e^d_{\beta}\Gamma^c_{\ ad}v
\gamma_c\gamma^b\gamma^a
+\frac{1}{4}\partial_a\delta e^c_{\beta}e_b^{\beta}v
\gamma_c\gamma^b\gamma^a\nonumber\\
&&
-\frac{1}{4}\delta e^c_{\gamma}e_a^{\alpha}e_b^{\beta}
\Gamma^{\gamma}_{\ \alpha\beta}v\gamma_c\gamma^b\gamma^a
-\frac{1}{4}\delta\Gamma^{\gamma}_{\ \alpha\beta}
e_a^{\alpha}e_b^{\beta}e^c_{\gamma}v\gamma_c\gamma^b\gamma^a.
\end{eqnarray}

We can perform an integration by parts as follows:
\begin{eqnarray}\label{var2}
\frac{1}{4}\partial_a\delta e^c_{\beta}e_b^{\beta}v
\gamma_c\gamma^b\gamma^a&=&
\frac{-i}{4}P_c(\delta e^c_{\beta}e_b^{\beta}v\gamma_c\gamma^b)
+\frac{i}{4}\delta e^c_{\beta}e_b^{\beta}P_c(v\gamma_c\gamma^b)
\nonumber\\
&&-\frac{1}{4}\delta e^c_{\beta}\partial_ae_b^{\beta}v\gamma_c
\gamma^b\gamma^a
-\frac{1}{4}\delta e^d_{\beta}e_b^{\beta}\Gamma^c_{\ ad}
v\gamma_c\gamma^b\gamma^a
+\frac{1}{4}\delta e^c_{\beta}e_d^{\beta}\Gamma^d_{\ ab}
v\gamma_c\gamma^b
\gamma^a\nonumber\\
&=&\frac{-i}{4}P_c(\delta e^c_{\beta}e_b^{\beta}v\gamma_c\gamma^b)
+\frac{i}{4}\delta e^c_{\beta}e_b^{\beta}(P_cv)\gamma_c\gamma^b
-\frac{1}{4}\delta e^c_{\beta}e_b^{\beta}\nabla_av
\left[\gamma_c\gamma^b,\gamma^a\right]\\
&&-\frac{1}{4}\delta e^c_{\beta}\partial_ae_b^{\beta}v\gamma_c
\gamma^b\gamma^a
+\frac{1}{4}\delta e_b^{\beta}e^d_{\beta}\Gamma^c_{\ ad}
v\gamma_c\gamma^b\gamma^a
+\frac{1}{4}\delta e^c_{\beta}e_d^{\beta}\Gamma^d_{\ ab}
v\gamma_c\gamma^b\gamma^a.\nonumber
\end{eqnarray}
Because $\left[\gamma_c\gamma^b,\gamma^a\right]
=\gamma_c\left\{\gamma^b,\gamma^a\right\}
-\left\{\gamma_c,\gamma^a\right\}\gamma^b
=2\eta^{ab}\gamma_c-2\delta^a_c\gamma^b$
and $e^c_{\beta}=g_{\mu\beta}\eta^{cd}e_d^{\mu}$ we can write:
\begin{eqnarray}\label{var3}
-\frac{1}{4}\delta e^c_{\beta}e_b^{\beta}\nabla_av
\left[\gamma_c\gamma^b,\gamma^a\right]&=&
-\frac{1}{2}\delta(g_{\mu\beta}\eta^{cd} e_d^{\mu})e_b^{\beta}
\eta^{ab}\nabla_av\gamma_c
+\frac{1}{2}\delta e^c_{\beta}e_b^{\beta}\nabla_cv\gamma^b
\nonumber\\
&=&
-\frac{1}{2}\delta g_{\mu\beta}\eta^{cd} e_d^{\mu}e_b^{\beta}
\eta^{ab}\nabla_av\gamma_c
-\delta e_d^{\mu}e^a_{\mu}\nabla_av\gamma^d\nonumber\\
&=&
\frac{1}{2}\delta g^{\alpha\beta}e^a_{\alpha}e^b_{\beta}
\nabla_av\gamma_b
-\delta e_a^{\alpha}e^d_{\alpha}\nabla_dv\gamma^a.
\end{eqnarray}
When substituting equations (\ref{var2}) and (\ref{var3}) in
(\ref{var1}) we can recombine the terms
\[
\frac{-1}{4}\delta e^c_{\beta}\partial_ae_b^{\beta}
v\gamma_c\gamma^b\gamma^a-\frac{1}{4}\delta e^c_{\gamma}e_a^{\alpha}
e_b^{\beta}\Gamma^{\gamma}_{\ \alpha\beta}v\gamma_c\gamma^b\gamma^a
=\frac{-1}{4}\delta e^c_{\gamma}e_d^{\gamma}\Gamma^d_{\ ab}v
\gamma_c\gamma^b\gamma^a
\]
to obtain
\begin{eqnarray}\label{var4}
\delta\dirop v&=&
\frac{-i}{4}P_c(\delta e^c_{\beta}e_b^{\beta}v\gamma_c\gamma^b)
+\frac{i}{4}\delta e^c_{\beta}e_b^{\beta}(P_cv)\gamma_c\gamma^b
+\frac{1}{2}\delta g^{\alpha\beta}e^a_{\alpha}e^b_{\beta}
\nabla_av\gamma_b\\
&&-\frac{1}{4}\delta\Gamma^{\gamma}_{\ \alpha\beta}
e_a^{\alpha}e_b^{\beta}e^c_{\gamma}v\gamma_c\gamma^b\gamma^a.\nonumber
\end{eqnarray}
Note that the variations of the frame $\delta e_a^{\alpha}$ cancel out,
except in the terms with $P_c$. These are harmless when we compute
$B_0(\delta \dirop S_0f)$, because both $B_0$ and $v$ solve the Dirac
equation. Therefore, the final answer will not depend on variations of
the frame, as desired.

In the last term of equation (\ref{var4}) we can use the
symmetry of the Christoffel symbol:
\begin{eqnarray}\label{var5}
-\frac{1}{4}\delta\Gamma^{\gamma}_{\ (\alpha\beta)}
e_a^{\alpha}e_b^{\beta}e^c_{\gamma}v\gamma_c\gamma^b\gamma^a&=&
-\frac{1}{4}\delta\Gamma^{\gamma}_{\ \alpha\beta}
e_a^{\alpha}e_b^{\beta}e^c_{\gamma}v\gamma_c\eta^{ab}
=-\frac{1}{4}\delta\Gamma^{\gamma}_{\ \alpha\beta}g^{\alpha\beta}
e^c_{\gamma}v\gamma_c\\
&=&
-\frac{1}{4}\delta g^{\gamma\mu}g_{\mu\nu}\Gamma^{\nu}_{\ \alpha\beta}
g^{\alpha\beta}e^c_{\gamma}v\gamma_c
-\frac{1}{4}\partial_{\alpha}\delta g_{\beta\mu}e_a^{\mu}
g^{\alpha\beta}v\gamma^a\nonumber\\
&&+\frac{1}{8}\partial_{\mu}\delta g_{\alpha\beta}e_a^{\mu}
g^{\alpha\beta}v\gamma^a\nonumber
\end{eqnarray}
We handle the last term using an integration by parts as before:
\begin{eqnarray}\label{var6}
\frac{1}{8}\partial_a\delta g_{\alpha\beta}g^{\alpha\beta}v\gamma^a
&=&\frac{-i}{8}P_c(\delta g_{\alpha\beta}g^{\alpha\beta}v)
+\frac{i}{8}\delta g_{\alpha\beta}g^{\alpha\beta}P_cv
-\frac{1}{8}\delta g_{\alpha\beta}\partial_ag^{\alpha\beta}v\gamma^a\\
&=&\frac{-i}{8}P_c(\delta g_{\alpha\beta}g^{\alpha\beta}v)
+\frac{i}{8}\delta g_{\alpha\beta}g^{\alpha\beta}P_cv
-\frac{1}{8}\delta g^{\alpha\beta}\partial_ag_{\alpha\beta}v
\gamma^a,\nonumber
\end{eqnarray}
where we used $\delta g_{\alpha\beta}\partial_ag^{\alpha\beta}=
-\delta g^{\alpha\beta}g_{\alpha\mu}g_{\beta\nu}\partial_ag^{\mu\nu}
=\delta g^{\alpha\beta}\partial_ag_{\alpha\beta}$.
The penultimate term in (\ref{var5}) is:
\begin{eqnarray}\label{var7}
-\frac{1}{4}\partial_{\alpha}\delta g_{\beta\mu}e_a^{\mu}
g^{\alpha\beta}v\gamma^a&=&
\frac{1}{4}\partial_b(\delta g^{\alpha\beta}g_{\alpha\mu}g_{\beta\nu})
e_a^{\mu}e^b_{\rho}g^{\rho\nu}v\gamma^a\nonumber\\
&=&\frac{1}{4}\partial_b(\delta g^{\alpha\beta}e^a_{\alpha}e^b_{\beta})
v\gamma_a-\frac{1}{4}\delta g^{\alpha\beta}g_{\alpha\mu}g_{\beta\nu}
\partial_b(e_a^{\mu}e^b_{\rho}g^{\nu\rho})v\gamma^a\nonumber\\
&=&
\frac{1}{4}\nabla_b(\delta g^{\alpha\beta}e^a_{\alpha}e^b_{\beta})
v\gamma_a
-\frac{1}{4}\delta g^{\alpha\beta}\left(\Gamma^a_{\ bc}e^c_{\alpha}
e^b_{\beta}+\Gamma^b_{\ bc}e^a_{\alpha}e^c_{\beta}\right)v\gamma_a
\nonumber\\
&&
-\frac{1}{4}\delta g^{\alpha\beta}g_{\alpha\mu}g_{\beta\nu}
\partial_b(e_a^{\mu}e^b_{\rho}g^{\nu\rho})v\gamma^a.
\end{eqnarray}
The first term on the right-hand side of equation (\ref{var7}) is
\begin{equation}\label{var8}
\frac{1}{4}\nabla_b(\delta g^{\alpha\beta}e^a_{\alpha}e^b_{\beta})
v\gamma_a=
\frac{1}{4}\nabla_b(\delta g^{\alpha\beta}e^a_{\alpha}e^b_{\beta}
v\gamma_a)-
\frac{1}{4}\delta g^{\alpha\beta}e^a_{\alpha}e^b_{\beta}
\nabla_bv\gamma_a.
\end{equation}
The other terms can be simplified with some computation:
\begin{eqnarray}\label{var9}
&&-\frac{1}{4}\delta g^{\alpha\beta}\left(\Gamma^a_{\ bc}e^c_{\alpha}
e^b_{\beta}+\Gamma^b_{\ bc}e^a_{\alpha}e^c_{\beta}+
g_{\alpha\mu}g_{\beta\nu}\eta^{ac}
\partial_b(e_c^{\mu}e^b_{\rho}g^{\rho\nu})\right)v\gamma_a
\nonumber\\
&=&-\frac{1}{4}\delta g^{\alpha\beta}\left(
-\partial_{\beta}e^a_{\alpha}
+e^a_{\gamma}\Gamma^{\gamma}_{\ \beta\alpha}
-e^a_{\alpha}\partial_ce^c_{\beta}
+e^a_{\alpha}\Gamma^{\mu}_{\ \mu\beta}
+e^a_{\alpha}g_{\beta\nu}\partial_{\rho}g^{\rho\nu}
+e^a_{\alpha}\partial_be^b_{\beta}
+g_{\alpha\mu}\eta^{ac}\partial_{\beta}e_c^{\mu}
\right)v\gamma_a\nonumber\\
&=&-\frac{1}{4}\delta g^{\alpha\beta}\left(
-\eta^{ac}e_c^{\mu}\partial_{\beta}g_{\alpha\mu}
+e^a_{\gamma}\Gamma^{\gamma}_{\ \beta\alpha}
+e^a_{\alpha}\Gamma^{\mu}_{\ \mu\beta}
-e^a_{\alpha}g^{\rho\nu}\partial_{\rho}g_{\beta\nu}
\right)v\gamma_a\nonumber\\
&=&-\frac{1}{8}\delta g^{\alpha\beta}\left(
-2e^a_{\gamma}g^{\gamma\mu}\partial_{\beta}g_{\alpha\mu}
+e^a_{\gamma}g^{\gamma\mu}(2\partial_{\beta}g_{\alpha\mu}
-\partial_{\mu}g_{\alpha\beta})
+e^a_{\alpha}g^{\mu\gamma}\partial_{\beta}g_{\mu\gamma}
-2e^a_{\alpha}g^{\rho\nu}\partial_{\rho}g_{\beta\nu}
\right)v\gamma_a\nonumber\\
&=&\frac{1}{8}\delta g^{\alpha\beta}\left(
e^a_{\gamma}g^{\gamma\mu}
\partial_{\mu}g_{\alpha\beta}
+2e^a_{\alpha}g_{\beta\mu}g^{\rho\nu}\Gamma^{\mu}_{\ \rho\nu}
\right)v\gamma_a.
\end{eqnarray}
Substituting equations (\ref{var5}-\ref{var9}) into (\ref{var4})
yields:
\begin{eqnarray}\label{var10}
\delta\dirop v&=&
\frac{-i}{4}P_c(\delta e^c_{\beta}e_b^{\beta}v\gamma_c\gamma^b)
+\frac{i}{4}\delta e^c_{\beta}e_b^{\beta}(P_cv)\gamma_c\gamma^b
-\frac{i}{8}P_c(\delta g_{\alpha\beta}g^{\alpha\beta}v)
+\frac{i}{8}\delta g_{\alpha\beta}g^{\alpha\beta}P_cv\nonumber\\
&&+\frac{1}{4}\delta g^{\alpha\beta}e^a_{\alpha}e^b_{\beta}
\nabla_av\gamma_b
+\frac{1}{4}\nabla_b(\delta g^{\alpha\beta}e^a_{\alpha}e^b_{\beta}
v\gamma_a).
\end{eqnarray}
Using Lemma \ref{adjointvar} we find for a spinor $u\in\Test(DM)$:
\begin{eqnarray}\label{var11}
\delta\dirop u&=&
\frac{i}{4}P(\delta e^c_{\beta}e_b^{\beta}\gamma^b\gamma_cu)
-\frac{i}{4}\delta e^c_{\beta}e_b^{\beta}\gamma^b\gamma_c(Pu)
+\frac{i}{8}P(\delta g_{\alpha\beta}g^{\alpha\beta}u)
-\frac{i}{8}\delta g_{\alpha\beta}g^{\alpha\beta}Pu\nonumber\\
&&+\frac{1}{4}\delta g^{\alpha\beta}e^a_{\alpha}e^b_{\beta}
\gamma_b\nabla_au
+\frac{1}{4}\nabla_b(\delta g^{\alpha\beta}e^a_{\alpha}e^b_{\beta}
\gamma_au).
\end{eqnarray}

Using Proposition \ref{varRCE} and equations (\ref{var10},\ref{var11})
we notice that the terms with $P_c$ and $P$ cancel out in the following
equality, because $B_0$ and $S_0f$ both satisfy the Dirac equation:
\begin{eqnarray}
\delta(\beta_{\epsilon}B_0(f))&=&
-B_0(\delta P_{\epsilon}S_0f)
=\frac{i}{4}B_0(\delta g^{\alpha\beta}e^a_{\alpha}e^b_{\beta}
\gamma_b\nabla_aS_0Rf)
+\frac{i}{4}B_0(\nabla_b(\delta g^{\alpha\beta}e^a_{\alpha}e^b_{\beta}
\gamma_aS_0Rf))\nonumber\\
&=&\frac{i}{4}\delta g^{\alpha\beta}e^a_{\alpha}e^b_{\beta}
\left(B_0(\gamma_{(b}\nabla_{a)}S_0Rf)-\nabla_{(b}B_0(\gamma_{a)}S_0Rf)
\right).
\end{eqnarray}
We now compare with Proposition \ref{SEMTresult} to get the final result.

\end{document}